\documentclass{elsarticle}
\usepackage[utf8]{inputenc} 
\usepackage{amsthm,amsmath,amssymb}  
\usepackage{bm}
\usepackage{comment}
\usepackage[normalem]{ulem}
\usepackage{relsize} 

\newcommand{\xx}{\mathbf{x}}
\newcommand{\XX}{\mathbf{X}}

\newcommand{\XXX}{\mathbb{X}}
\newcommand{\XXXX}{\boldsymbol{\mathbb{X}}}
\newcommand{\YYYY}{\boldsymbol{\mathbb{Y}}}
\newcommand{\diff}{\mathrm{d}}
\newcommand{\Id}{\mathbf{I}}
\newcommand{\Tr}{\mathrm{Tr}}
\newcommand{\CC}{\mathbf{C}}
\newcommand{\dX}{\mathrm{d}\XX}
\newcommand{\dx}{\mathrm{d}\xx}
\newcommand{\mm}{\mathbf{m}}
\newcommand{\vv}{\mathbf{v}}
\newcommand{\MM}{\mathbf{M}}
\newcommand{\OBig}{\mathcal{O}}
\newcommand{\eps}{\varepsilon}
\newcommand{\FF}{\mathbf{F}}
\renewcommand{\AA}{\mathbf{A}}
\newcommand{\BB}{\mathbf{B}}
\newcommand{\DD}{\mathbf{D}}
\newcommand{\qq}{\mathbf{q}}
\newcommand{\rr}{\mathbf{r}}
\newcommand{\cc}{\mathbf{c}}
\newcommand{\pp}{\mathbf{p}}
\newcommand{\ww}{\mathbf{w}}
\newcommand{\WW}{\mathbf{W}}
\newcommand{\QQ}{\mathbf{Q}}
\newcommand{\LL}{\mathbf{L}}
\newcommand{\Lie}{\mathfrak{L}}

\newcommand{\IP}[1]{{\color{Red}IP:\ \ #1}}

\newcommand{\sA}{{\mathsmaller A}}
\newcommand{\sB}{{\mathsmaller B}}

\newcommand{\sL}{{\mathsmaller L}}
\newcommand{\sK}{{\mathsmaller K}}
\newcommand{\sI}{{\mathsmaller I}}
\newcommand{\sJ}{{\mathsmaller J}}

\newcommand{\pd}{\partial}
\newcommand{\F}[2]{F^{#1}_{\ \, \mathsmaller#2}}
\newcommand{\hatF}[2]{\hat{F}^{#1}_{\ \, \mathsmaller#2}}
\newcommand{\A}[2]{A^{\mathsmaller#1}_{\ \, #2}}
\newcommand{\vab}[2]{v^{#1}_{\ \, #2}}
\newcommand{\curl}{\mathrm{curl}}
\newcommand{\Ffunc}{F}
\newcommand{\Gfunc}{G}
\newcommand{\Hfunc}{H}

\newcommand{\LA}{\mathfrak{l}}
\newcommand{\TRT}{\mathrm{TRT}}
\newcommand{\Par}{\mathcal{P}}

\usepackage[colorlinks]{hyperref}
\hypersetup{
	colorlinks=true, breaklinks=true, bookmarks=true,bookmarksnumbered,
	urlcolor=webbrown, linkcolor=RoyalBlue, citecolor=webgreen, 
	pdftitle={}, 
	pdfauthor={\textcopyright}, 
	pdfsubject={}, 
	pdfkeywords={}, 
	pdfcreator={pdfLaTeX}, 
	pdfproducer={LaTeX with hyperref and ClassicThesis} 
}

 \newtheorem{theorem}{Theorem}[section]
 \newtheorem{prop}[theorem]{Proposition}
 \theoremstyle{definition}
 \newtheorem{definition}[theorem]{Definition}
 \theoremstyle{remark}
 \newtheorem{remark}[theorem]{Remark}
 
 \numberwithin{equation}{section}

\begin{document}

\title{On Hamiltonian continuum mechanics}
\author[mff]{Michal Pavelka\corref{cor1}}
\ead{pavelka@karlin.mff.cuni.cz}
\cortext[cor1]{Corresponding author}
\author[Tre]{Ilya Peshkov}
\author[fjfi]{Václav Klika}
\address[mff]{Mathematical Institute, Faculty of Mathematics and Physics, Charles University, Sokolovská 83, 186 75 Prague, Czech Republic}
\address[Tre]{Laboratory of Applied Mathematics, Department of Civil, Environmental and Mechanical Engineering, University of Trento, Via Mesiano 77, 38123 Trento, Italy}
\address[fjfi]{Department of Mathematics, FNSPE, Czech Technical University in Prague, Trojanova 13, Prague 2, 120 00, Czech Republic}

\begin{abstract}

Continuum mechanics can be formulated in the Lagrangian frame (addressing motion of individual continuum particles) or in the Eulerian frame (addressing evolution of fields in an inertial frame). 
There is a canonical Hamiltonian structure in the 
Lagrangian frame. By transformation to the Eulerian frame we find the Poisson 
bracket for Eulerian continuum mechanics with deformation gradient (or the 
related distortion matrix). Both Lagrangian and Eulerian Hamiltonian structures 
are then discussed from the perspective of space-time variational formulation 
and by means of semidirect products and Lie algebras. Finally, we discuss the 
importance of the Jacobi identity in continuum mechanics and approaches to prove hyperbolicity 
of the evolution equations and their gauge invariance.
\end{abstract}


\maketitle

\numberwithin{equation}{section}


\textit{"Le savant n'étudie pas la nature parce que cela est utile; il l'étudie parce qu'il y prend plaisir et il y prend plaisir parce qu'elle est belle."}
$\qquad$ Henri Poincaré \cite{Poincare-science}.


\section{Introduction}\label{sec.intro}
The usual way continuum mechanics is presented is based on a generalization of Newton's laws to 
continuum particles \cite{BookGurtin,Marsden-hughes,Simo1988}. As Newton's laws can be seen as a consequence 
of the principle of least action or Hamiltonian mechanics (Hamilton canonical equations), see e.g. \cite{Landau1}, so can be 
the continuum mechanics. Moreover, continuum mechanics can be formulated in the Lagrangian frame, 
where coordinates are attached to matter, or Eulerian frame, where coordinates are attached to an 
inertial frame of reference. We shall present the novel reduction of the Hamiltonian continuum 
mechanics in the Lagrangian frame to the Eulerian frame. This leads to the Poisson bracket 
generating reversible evolution equations for density, momentum density and entropy density 
(balance laws) coupled with evolution for the deformation gradient or its inverse, which is called 
the distortion. 
This is a Hamiltonian formulation of 
continuum mechanics suitable for both solids and fluids\footnote{In this paper, when we talk about 
unified treatment of fluids and solids, by the term ``fluid'' we refer not to the \emph{ideal} 
fluids, which do not require any extra degrees of freedom related to a deformation measure, but to 
more general viscous fluids which in the discussed framework can be indeed considered 
as a particular case of inelasticity, e.g. \cite{HPR2016,DPRZ2016,HYP2016,Busto2019}.
} in the Eulerian frame.

We also discuss the importance of the \emph{hyperbolicity} \cite{Gavrilyuk-waves,Despres-book,Torrilhon-simplified} (or 
\emph{evolutionarity} in terms of Beris and Edwards \cite{BE}) for the time-dependent partial 
differential equations. In particular, we discuss the SHTC (Symmetric Hyperbolic Thermodynamically 
Compatible) equations for 
nonlinear Eulerian elasticity and fluid mechanics\footnote{Originally proposed by Godunov, 	
Romenskii et al. \cite{Godunov-interesting,God-Siberian}}. Recall that the hyperbolicity of a 
time-dependent PDE system 
involving derivatives up to the order $ N $ is a 
natural physical requirement which states that, at least locally in time, the solution (as well as 
its $ 
N-1 $ derivatives) to the initial value problem for the system exists, is unique and 
depends continuously on initial 
data (at $ t=0 $) along arbitrary non-characteristic hypersurfaces.
Our motivation here is to shed some light on the interconnections between the hyperbolicity of 
evolutionary PDEs and their Hamiltonian formulation which to the best of our knowledge has been not 
rigorously discussed in the literature. Recall the importance of the hyperbolicity 
also for the numerical resolution of evolutionary equations which guaranties that the small 
perturbations to the solution due to the discretization errors do not amplify exponentially in 
time.

Then, we turn to some interesting geometric features of Hamiltonian continuum mechanics.
The meaning of Jacobi identity is discussed in Sec.\,\ref{sec.Jacobi}, and a new relation between 
Hamiltonian mechanics and hyperbolicity of the 
governing evolution equations is developed in Sec.\,\ref{sec.hype}. In Sec.\,\ref{sec.gauge}, we 
show 
in a simplified way how existence 
of a conserved quantity implies gauge invariance of the evolution equations and vice versa
(even in the non-canonical case). In Sec.\,\ref{sec.Clebsch}, a new construction from 
the Clebsch variables 
(providing a variational principle) of fluid mechanics with the distortion field is presented. 
In Sec.\,\ref{sec.SP}, the Poisson bracket for Eulerian continuum mechanics with 
distortion field is 
shown to have the structure of semidirect product of fluid mechanics and a cotangent bundle, which 
provides new insight into the structure of the SHTC equations.

In Sec.\,\ref{sec.Lagr.mech}, we identify the variational structure of 
Lagrangian continuum mechanics in the four-dimensional space-time settings and the complementary 
Eulerian 
variational principle. The motivation for this section lies in the lack of a conventional 
four-dimensional formulation of the Hamiltonian continuum mechanics including distortion in which the time and space 
coordinates are not explicitly separated. The importance of such a formulation can be viewed not 
only from the theoretical standpoint of formulating relativistic continuum theories 
\cite{Ottinger1998,PTRSA2019}, but also from a very practical standpoint of designing 
structure-preserving\footnote{Such as constrain preserving methods (divergence-preserving in 
magnetohydrodynamics or curl-preserving in elasticity), energy preserving methods, and etc.} 
numerical methods (so-called symplectic integrators, variational integrators, etc.), see e.g. 
\cite{Morrison2017,Ottinger2018a,Gross2018}. Thus, our results 
demonstrate that the four-dimensional formalism allows to see that the equations we study, in 
fact, possess a more general space-time structure, in which the time is not separated from the 
space and that, in a structure-preserving numerical method, it should be treated in the same way as 
the space 
coordinates (e.g. with staggering in both space and time \cite{Ghrist2001}). Note that the choice 
of the time-integrator is very critical in such methods and the wrong choice may ruin the overall 
structure compatibility of the numerical solution, see e.g. \cite{ehrenfest-regularization} for a general Poisson integrator. 
Another motivation for the space-time variational principle is to understand the origins of 
previous results on relativistic formulation of the Hamiltonian continuum mechanics 
 \cite{Ottinger1998,HCO} without relying on the intuition gained in Galilean physics (time separated from spac).


Novelty of this paper lies (i) in the transformation from the Hamiltonian 
mechanics in the Lagrangian frame to the Eulerian frame (extending earlier results), (ii) in the 
new proof of hyperbolicity of a system of conservation laws and establishing connections between
hyperbolicity of the system and its Hamiltonian formulation, (iii) in the 
simplified approach to gauge invariance, 
(iv) in the semidirect-product 
structure of SHTC equations, (v) and in the space-time variational formulation 
of the equations.

\section{Hamiltonian mechanics}\label{sec.Ham}
The principle of least action has long been the fundamental approach to modern mechanics 
\cite{Landau1}. Although initially formulated for mechanics of classical particles and optics, 
 it has been applied also to the mechanics of rigid bodies, field 
theories 
and continuum mechanics, e.g. \cite{Arnold}. In the 80s, Hamiltonian mechanics was connected 
with thermodynamics in several related formulations \cite{dv,Morrison-brackets, g1, 
EdwardsBeris91,BE}, in particular in the GENERIC (General Equation for Non-Equilibrium 
Reversible-Irreversible Coupling) framework \cite{Grmela1997,Ottinger1997}, which has been covered 
in monographies \cite{HCO,PKG}.

The usual obstacle, however, for understanding the GENERIC framework is the presence of Poisson brackets, which are often regarded as a mysterious mathematical concept. We would like to elucidate their origin by demonstrating a simple-in-principle derivation of Hamiltonian Eulerian continuum mechanics requiring only minimal geometric background.

\subsection{Lagrangian frame}\label{sec.Lagrange}
Let us consider a body, material points of which are described by a Lagrangian (reference) 
coordinates $\XX$. Material points are elements of a material manifold. Position of the material point $\XX$ at time $t$ with respect to a chosen 
inertial laboratory frame is then given by the mapping $\xx(t,\XX)$ from the Lagrangian coordinates 
to the Eulerian coordinates. This mapping is usually assumed to be smooth enough and invertible. 
These properties will be violated later in this paper, but for the moment let us adopt those 
assumptions as well. 

Mechanical state of a material point is characterized by its position $\xx(t,\XX)$ and velocity $\dot{\xx}(t,\XX)$ or the corresponding momentum density $\MM(t,\XX)$ (momentum per Lagrangian volume $\diff \XX$). In mathematical terms the couple $(\xx,\dot{\xx})$ forms a tangent bundle while the couple $(\xx,\MM)$ forms a cotangent bundle. Since we are seeking Hamiltonian evolution (generated by a Poisson bracket and energy), we choose the latter description. The Lagrangian state variables are thus the field of Eulerian positions $\xx(t,\XX)$ and the field of momentum density $\MM(t,\XX)$. 

Since these state variables form a cotangent bundle, they are equipped with the \textit{canonical} 
Poisson bracket in the Lagrangian frame, see e.g. \cite{Simo1988},  
\begin{equation}\label{eq.PB.L}
	\{\Ffunc,\Gfunc\}^{(L)} = \int\dX \left(\frac{\delta \Ffunc}{\delta 
	x^i(\XX)} \frac{\delta \Gfunc}{\delta M_i(\XX)} -\frac{\delta 
	\Gfunc}{\delta x^i(\XX)} \frac{\delta \Ffunc}{\delta M_i(\XX)}\right),
\end{equation}
where $\Ffunc$ and $\Gfunc$ are two arbitrary functionals of the Lagrangian 
state variables. The explicit dependence on time is omitted from the notation 
as in the rest of the paper since now on. The derivatives stand for functional 
(or Volterra) derivatives, see \ref{sec.FD}. This Poisson bracket 
clearly satisfies the Jacobi identity,
\begin{subequations}\label{eq.PB.prop}
	\begin{equation}\label{eq.Jacobi}
	\{\Ffunc,\{\Gfunc,\Hfunc\}\}+
	\{\Gfunc,\{\Hfunc,\Ffunc\}\}+
	\{\Hfunc,\{\Ffunc,\Gfunc\}\} = 0,
\end{equation}
as can be seen by direct verification. The bracket is of course antisymmetric,
\begin{equation}
	\{\Ffunc,\Gfunc\} = -\{\Gfunc,\Ffunc\},
\end{equation}
and satisfies the Leibniz rule (assuming sufficient mathematical regularity),
\begin{equation}
	\{\Ffunc,\Gfunc \Hfunc\} = \{\Ffunc,\Gfunc\}\Hfunc + 
	\Gfunc\{\Ffunc,\Hfunc\}.
\end{equation}
\end{subequations}
Therefore, bracket \eqref{eq.PB.L} is indeed a Poisson bracket as it satisfies all the properties \eqref{eq.PB.prop}. 

Denoting a general set of state variables by $\qq$, a Poisson bracket can be equivalently expressed by means of its Poisson bivector\footnote{Greek indexes denote state variables while Latin space (or space-time) coordinates.}
\begin{equation}\label{bivector}
	L^{\alpha \beta} = \{q^\alpha,q^\beta\},
\end{equation}
which is antisymmetric and can be used to reconstruct the bracket as follows,
\begin{equation}
	\{\Ffunc,\Gfunc\} = \langle \Ffunc_{q^\alpha}| L^{\alpha\beta}| \Gfunc_{q^\beta}\rangle,
\end{equation}
where $\langle\bullet|\bullet\rangle$ means a contraction (e.g. integration over space or duality in distributions) and we shall use Dirac bra-ket notation below.

Once having some state variables $\qq$, e.g. $\qq=(\xx(\XX),\MM(\XX))$ in the Lagrangian continuum mechanics, and the 
corresponding Poisson bracket, the reversible evolution of a functional 
$\Ffunc(\qq)$ of the state variables is given by 
\begin{equation}
	\dot{\Ffunc} = \{\Ffunc,E\},
\end{equation}
where $E$ is the total energy of the (isolated) system. This is a sort of weak formulation of the problem. On the other hand, evolution of the functional $F$ can be expressed using the chain rule as functional derivatives of the functional multiplied by evolution equations of the state variables,
\begin{equation}\label{eq.qL}
	\dot{q}^\alpha = \{q^\alpha, E\} = \left. L^{\alpha\beta}\Big| \frac{\delta E}{\delta q^\beta}\right\rangle.
\end{equation}
For instance, for the Lagrangian state variables we have 
\begin{equation}
	\dot{\Ffunc}(\xx(\XX),\MM(\XX)) = \{\Ffunc,E\}^{(L)}
\end{equation}
as well as
\begin{equation}
	\dot{\Ffunc}(\xx(\XX),\MM(\XX)) = \int\dX \left(\frac{\delta \Ffunc}{\delta 
	x^i} \partial_t x^i + \frac{\delta \Ffunc}{\delta M_i} \partial_t M_i 
	\right).
\end{equation}
By comparing these two equalities, we can conclude that the evolution equations for $\xx$ and $\MM$ are
\begin{subequations}\label{eq.L.evo}
	\begin{eqnarray}
		\partial_t x^i(\XX) &=& \frac{\delta E}{\delta M_i(\XX)}\\
		\partial_t M_i(\XX) &=& -\frac{\delta E}{\delta x^i(\XX)}\label{eq.L.evo.mom}
	\end{eqnarray}
\end{subequations}
	for any energy $E(\xx,\MM)$. 
This is a way to obtain evolution equations from a Poisson bracket.

Let us be more specific. Choosing the energy as
\begin{equation}\label{eq.Ene.L}
	E = \int\dX \left(\frac{\MM^2}{2\rho_0} + \rho_0 W(\nabla_\XX \xx)\right),
\end{equation}
where the first term denotes the kinetic energy and the second denotes elastic energy (dependent only on gradients of the field $\xx(\XX)$), equations \eqref{eq.L.evo} obtain the concrete form
\begin{subequations}\label{eq.L.evo.fin}
	\begin{eqnarray}
		\partial_t x^i(\XX) &=& \frac{\delta^{ij} M_j}{\rho_0}\\
		\partial_t M_i(\XX) &=& \frac{\partial}{\partial X^I}\left(\rho_0 \frac{\partial W}{\partial \frac{\partial x^i}{\partial X^I}}\right)
	\end{eqnarray}
\end{subequations}
Here, $\rho_0(\XX)$ is a reference mass density field. The metric tensor $\delta^{ij}$ can be thought of as equal to the 
	unit matrix in the Euclidean space endowed with Cartesian coordinates. Note that the Einstein 
	summation convention is employed and that the capital index denotes coordinates in the 
	Lagrangian frame. Also, apart form the field $\rho_0(\XX)$ the energy can depend on the field of 
	entropy density $s_0(\XX)$ (per volume $\dX$) to cope with non-isothermal bodies. Equations 
	\eqref{eq.L.evo.fin} are the reversible evolution equations for a continuous body with stored 
	energy $W(\nabla_\XX \xx)$ in the Lagrangian frame, which are to be solved when initial and 
	boundary conditions are supplied. 

However, it is often preferable to formulate the evolution equations in the Eulerian frame because (i) the Lagrangian configuration may be inaccessible (as in the case of fluids), (ii) conservation laws are directly at hand in the Eulerian frame and so it is clearer how to add dissipative terms to the evolution equations, and (iii) also, the use of the Eulerian formulation can be advantageous in many practical situations \cite{Hyper-Hypo2019}. The complementary equations in the Eulerian frame are shown in the next section.

\subsection{Eulerian frame}\label{sec.Euler}
First we have to declare what are the fields constituting the Eulerian state variables. We choose the fields
\begin{subequations}\label{eq.x.E}
	\begin{eqnarray}
		\rho(\xx) &=& \rho_0(\XX(\xx)) \cdot \det \frac{\partial \XX}{\partial \xx}\\
		\mm(\xx) &=& \MM(\XX(\xx)) \cdot \det \frac{\partial \XX}{\partial \xx}\\
		s(\xx) &=& s_0(\XX(\xx)) \cdot \det \frac{\partial \XX}{\partial \xx}\\
		\F{i}{I}(\xx) &=& \frac{\partial x^i}{\partial 
		X^I}\Big|_{\XX(\xx)}\label{state.var.Euler.def}
	\end{eqnarray}
\end{subequations}
of local mass density (per volume $\dx$), momentum density, entropy density and the deformation 
gradient.

The goal is to project the Lagrangian Poisson bracket \eqref{eq.PB.L} to an Eulerian Poisson bracket by letting the functionals depend only on the Eulerian fields \eqref{eq.x.E}. After rather lengthy but straightforward calculation (\ref{sec.L-E})\footnote{A simpler version of the calculation leading to fluid mechanics was called a "small miracle" in \cite{Abarbanel}, and similar procedure leading to fluid mechanics equipped with the left Cauchy-Green tensor was presented in \cite{BE}.}, we obtain the Poisson bracket
\begin{eqnarray}\label{eq.PB.Eu}
	\{\Ffunc,\Gfunc\}^{(Eulerian)} &=& \{\Ffunc,\Gfunc\}^{(FM)} + \int\dx 
	\F{j}{I} 
	\left(\frac{\delta \Ffunc}{\delta 
	\F{i}{I}} \partial_j \frac{\delta \Gfunc}{\delta m_i}-\frac{\delta 
	\Gfunc}{\delta \F{i}{I}} \partial_j 
	\frac{\delta \Ffunc}{\delta m_i}\right)\nonumber\\
	&&-\int\dx \partial_k \F{i}{I} \left(\frac{\delta \Ffunc}{\delta \F{i}{I}} 
	\frac{\delta \Gfunc}{\delta 
	m_k}-\frac{\delta \Gfunc}{\delta \F{i}{I}} \frac{\delta \Ffunc}{\delta 
	m_k}\right),
\end{eqnarray}
where 
$\{\Ffunc,G\}^{(FM)}$ stands for the Poisson bracket of fluid mechanics,
\begin{multline}\label{eq.PB.FM}
	\{\Ffunc,\Gfunc\}^{(FM)} = \int\dx \rho (\partial_i \Ffunc_\rho 
	\Gfunc_{m_i}-\partial_i \Gfunc_\rho \Ffunc_{m_i})\\
	+ \int\dx m_i (\partial_j \Ffunc_{m_i} \Gfunc_{m_j}-\partial_j \Gfunc_{m_i} 
	\Ffunc_{m_j})\\
	+ \int\dx s (\partial_i \Ffunc_s \Gfunc_{m_i}-\partial_i \Gfunc_s 
	\Ffunc_{m_i}).
\end{multline}
For brevity, from now on, the functional derivatives in the Poisson brackets 
are denoted by subscript, e.g. $\frac{\delta \Ffunc}{\delta \rho} = 
\Ffunc_\rho$, 
and if the functionals are assumed to be local (involving no spatial 
gradients), $\Ffunc_\rho$ stands for the partial derivative $\Ffunc_\rho = \pd 
f/\pd 
\rho$, $f$ being volume density of $\Ffunc$. This slightly overloaded notation 
helps 
to keep the formulas clear and should not cause any confusion.
Bracket \eqref{eq.PB.Eu} is certainly a Poisson bracket, i.e. it fulfills criteria \eqref{eq.PB.prop}, since it has been obtained by reduction of Poisson brackets.\footnote{In \cite{PhysD-hierarchy} the reduction is called projection because the space of functionals of state variables is projected to the subspace of functionals dependent only on a submanifold of the state variables. If, after plugging in the functionals dependent only on the reduced state variables, the bracket depends only on the reduced state variables, i.e. is in a closed form, then it inherits the Jacobi identity from the original bracket and is indeed a Poisson bracket.} The bracket is compatible with Poisson bivector 5.12a of \cite{Markus2009}, and it expresses kinematics of the Eulerian state variables consisting of the state variables of fluid mechanics ($\rho$, $\mm$ and $s$) and the deformation gradient $\FF(\xx)$. The reduction can be summarized as the following theorem:
\begin{theorem}
Canonical Poisson bracket \eqref{eq.PB.L} of functionals $\Ffunc$ and $\Gfunc$ 
of the Eulerian fields \eqref{eq.x.E}  is equal 
to Poisson bracket \eqref{eq.PB.Eu}.
The latter bracket expresses kinematics of the Eulerian fields $(\rho,\mm,s,\FF)$.
\end{theorem}

Note that Poisson bracket \eqref{eq.PB.Eu} is more general than the bracket of fluid mechanics with 
the left Cauchy-Green tensor derived in \cite{BE} because the latter can be obtained from the 
former by a mapping, see Sec. \ref{sec.NN}, but not vice versa. On the other hand, bracket \eqref{eq.PB.Eu} is equivalent to the Poisson bracket (11)  from \cite{Hutter2008a} although the latter is formulated in momentum, temperature (instead of entropy) and deformation tensor (thus related by a one-to-one transformation). 

Bracket \eqref{eq.PB.Eu} is indeed a Poisson bracket, i.e. satisfies antisymmetry, Leibniz rule and Jacobi identity. The latter property is usually difficult to check (see \cite{kroeger2010} for an automated verification method). However, if the reduction ends up in a closed form, i.e. everything depends only on the reduced state variables, Jacobi identity is automatically satisfied as shown in \cite{PhysD-hierarchy}.

The reversible evolution equations implied by bracket \eqref{eq.PB.Eu} are
\begin{subequations}\label{eq.evo.Eu}
\begin{eqnarray}
	\partial_t \rho &=& -\partial_i(\rho E_{m_i})\\
	\partial_t m_i &=& -\partial_j(m_i E_{m_j}) -\rho\partial_i E_\rho - m_j \partial_i E_{m_j} -s 
	\partial_i E_s - \F{j}{J}\partial_i E_{\F{j}{J}} \nonumber\\
	&&+\partial_j(\F{j}{I} E_{\F{i}{I}} + \F{i}{I} E_{\F{i}{I}})\\
	\partial_t s &=& -\partial_i (s E_{m_i})\\
	\partial_t \F{i}{I} &=& -E_{m_k}\partial_k \F{i}{I} +  \F{j}{I}\partial_j 
	E_{m_i},\label{eq.evo.Eu.F}
\end{eqnarray}
	where the energy $E=\int\dx e(\rho,\mm,s,\FF)$ still remains to be specified. The functional derivatives of energy can be replaced by partial derivatives of total energy density $e$ hereafter due to the algebraic dependence on the state variables. 
	Note that the total momentum is 
        conserved, since the first line (except the first term) in the evolution equation for $m_i$ can be rewritten as gradient of generalized pressure $\partial_i p$ for 
	\begin{equation}\label{eq.p}
		p = -e + \rho \frac{\partial e}{\partial \rho} + m_j \frac{\partial e}{\partial m_j}+ 
		s\frac{\partial e}{\partial s} + \F{i}{I} \frac{\partial e}{\partial \F{i}{I}}.
	\end{equation}
\end{subequations}
Total energy density $e$ can be prescribed as
\begin{equation}\label{eq.ene.Eu}
	e = \frac{\mm^2}{2\rho} + \eps(\rho,s,\FF),
\end{equation}
where $\eps$ is the elastic and internal energy. In particular, $E_\mm = \mm/\rho = \vv$ becomes the velocity. The evolution equation for the deformation gradient then gets the explicit form
\begin{equation}
	\partial_t \FF = -(\vv\cdot\nabla) \FF + \nabla \vv \cdot \FF,
\end{equation}
which is the usual evolution equation for $\FF$ in the Eulerian frame, e.g. see \cite{GodRom2003,Hyper-Hypo2019,Rubin1987}. Equations \eqref{eq.evo.Eu} represent evolution equations for density, momentum density, entropy density and deformation gradient in the Eulerian frame, and they attain an explicit form once total energy density is specified.

Note that $ \FF $ 
(or of its inverse $ \AA $) can be generalized from being a holonomic triad (i.e. being the gradient 
of 
the mapping $ x^i(X^\sI) $) to non-holonomic triad. This, however, requires the introduction of the local 
reference configuration instead of the global Lagrangian configuration 
associated with the coordinates $ X^\sI $, e.g. see \cite{PRD-Torsion2019}.

\subsection{Non-Newtonian fluids}\label{sec.NN}


Since the Lagrangian configuration is usually irrelevant in the case of fluids (even non-Newtonian), the fluids are often described by state variables $\rho$, $\mm$, $s$ and the left Cauchy-Green tensor
\begin{equation}
	B^{ij}(\xx) = \F{i}{I}(\xx) \F{j}{J}(\xx) \delta^{\sI\sJ}.
\end{equation}
Note that $\delta^{\sI\sJ}$ is actually an inverse body metric measuring 
lengths in the Lagrangian frame, see \cite{Simo1988,Tamas-kinematics}.
The mapping from $\FF$ to $\BB$ is non-invertible, since $\FF$ has nine independent components 
while the symmetric tensor $\BB$ only six. Therefore, we shall refer to it as a reduction.
By 
letting the functionals in bracket \eqref{eq.PB.Eu} depend on these state 
variables we arrive at Poisson bracket
\begin{eqnarray}\label{eq.PB.B}
	\{\Ffunc,\Gfunc\}^{(LCG)} &=& \{\Ffunc,\Gfunc\}^{(FM)} \nonumber\\
	&&+ \int\dx \left(\Ffunc_{B^{ik}}(B^{jk}\partial_j 
	\Gfunc_{m_i}+B^{ji}\partial_j 
	\Gfunc_{m_k})-\Gfunc_{B^{ik}}(B^{jk}\partial_j \Gfunc_{m_i}+B^{ji}\partial_j
	 \Ffunc_{m_k})\right)\nonumber\\
	&&-\int\dx \partial_j 
	B^{ik}(\Ffunc_{B^{ik}}\Gfunc_{m_j}-\Gfunc_{B^{ik}}\Ffunc_{m_j}),
\end{eqnarray}
which expresses kinematics of fields $\rho$, $\mm$, $s$ and $\BB$.
Details of the calculation can be found in \ref{sec.F-B}. The description of a fluid including the left Cauchy-Green tensor is suitable for non-Newtonian complex fluids, e.g. \cite{Malek-Maxwell}. One can, however, describe non-Newtonian fluids also on the level involving  
deformation gradient (i.e. without the further reduction to $\BB$) \cite{Jackson2018}.

The evolution equations generated by bracket \eqref{eq.PB.B} are
\begin{subequations}
	\begin{eqnarray}
	\partial_t \rho &=& -\partial_i(\rho E_{m_i})\\
	\partial_t m_i &=& -\partial_j(m_i E_{m_j})-\rho\partial_i E_\rho - m_j \partial_i E_{m_j} -s \partial_i E_s - B^{jk}\partial_i E_{B^{jk}} \nonumber\\
	&&+\partial_i(B^{jk}E_{B^{jk}}) + \partial_j(B^{jk}(E_{B^{ik}}+E_{B^{ki}}))\\
	\label{eq.B.evo}\partial_t B^{ik} &=& -v^j \partial_j B^{ik} + B^{jk}\partial_j E_{m_i} + B^{ji}\partial_j E_{m_k}\\
	\partial_t s &=& -\partial_i(s E_{m_i}).
	\end{eqnarray}
\end{subequations}
The equation for the left Cauchy-Green tensor can be rewritten as 
$\stackrel{\nabla}{\BB}=0$, i.e. the upper-convected derivative of $\BB$ be 
equal to zero.\footnote{Note that we assume that kinetic energy is in form $\mm^2/2\rho$ so that its conjugate is velocity, $E_\mm = \vv$.}
Moreover, once the dependence of energy on the state variables is specified, 
the stress is determined and the equations get an explicit form, see e.g. 
\cite{PKG}.

Note that by letting the functionals depend only on the state variables of fluid mechanics, 
$\rho$, $\mm$ and $s$, Poisson bracket \eqref{eq.PB.B} is projected to the Poisson bracket of fluid mechanics \eqref{eq.PB.FM}. 
The $\{\bullet,\bullet\}^{(FM)}$ bracket then leads to the compressible Euler equations, but when the energy is chosen quadratic in gradients of density, one automatically gets equations for Korteweg fluid equations, see e.g. \cite{PKG} for more details.
Fluid mechanics can be also seen as evolution on the coadjoint orbit of the infinite-dimensional 
Lie group of diffeomorphism of a domain to itself \cite{marsden1984semidirect,Arnold}. The Hamiltonian formulation of fluid mechanics is especially useful for instance in analysis of stability and robustness \cite{Abarbanel-stability, Holm-balance}.

In summary, dynamics with the left Cauchy-Green tensor is less detailed than 
dynamics with deformation tensor or distortion, since the former is obtained by reduction of the 
latter ($\BB$ has six independent components while $\FF$ nine), and knowledge 
of $\FF$ allows reconstruction of the field of labels $\XX(\xx)$ by contour 
integration. 

\subsection{SHTC equations}\label{sec.SHTC}

Besides the reduction 
from $\FF$ to $\BB$, one can also carry out transformation of variables from 
$\FF$ to the distortion $\AA = \FF^{-1}$,
\begin{equation}
	\A{I}{i}(\xx) = (\FF^{-1}(\xx))^\sI_{\ i},
\end{equation}
by letting the functionals depend only on $\rho$, $\mm$, $s$ and $\AA$, see 
\ref{sec.F-A} for details. 

The resulting Poisson bracket is
\begin{eqnarray}\label{eq.PB.A}
	\{\Ffunc,\Gfunc\}^{(A)} &=& \{\Ffunc,\Gfunc\}^{(FM)} - \int\dx \A{L}{i} 
	(\Ffunc_{\A{L}{l}} \partial_l 
	\Gfunc_{m_i}-\Gfunc_{\A{L}{l} } \partial_l \Ffunc_{m_i})\nonumber\\
	&&-\int\dx \partial_i \A{L}{l} 
	(\Ffunc_{\A{L}{l}}\Gfunc_{m_i}-\Gfunc_{\A{L}{l}}\Ffunc_{m_i}),
\end{eqnarray}
which is the Poisson bracket for the distortion. Thus we come to the following conclusion, which might be anticipated already from the results in \cite{SHTC-GENERIC}.
\begin{prop}[On the origin of continuum mechanics with distortion]
The Poisson bracket \eqref{eq.PB.A}, which expresses kinematics of the Eulerian fields $(\rho,\mm,s,\AA)$, is obtained by reduction of the canonical Lagrangian Poisson bracket \eqref{eq.PB.L}.
\end{prop}

The reversible evolution equations generated by this Poisson bracket are
\begin{subequations}\label{eq.evo.A}
	\begin{eqnarray}
	\partial_t \rho &=& -\partial_i(\rho E_{m_i})\\
	\partial_t m_i &=& -\partial_j(m_i E_{m_j})-\rho\partial_i E_\rho - m_j \partial_i E_{m_j} -s 
	\partial_i E_s - \A{L}{l}\partial_i E_{\A{L}{l}} \nonumber\\
		&&+\partial_i(\A{L}{l} E_{\A{L}{l}}) - \partial_l(\A{L}{i} E_{\A{L}{l}})\\
	\partial_t s &=& -\partial_i(s E_{m_i}),\\
		\partial_t \A{L}{l} &=& -\partial_l (\A{L}{i} E_{m_i}) + (\partial_l \A{L}{i} - \partial_i 
		\A{L}{l}) 
		E_{m_i}.\label{eq.evo.Eu.A}
	\end{eqnarray}
\end{subequations}
Again, once the energy is specified, the equations acquire an explicit form. These evolution 
equations are part of the Symmetric Hyperbolic Thermodynamically Compatible (SHTC) equations, 
originally found in \cite{God1961}, see also \cite{Godunov1996,SHTC-GENERIC, 
PKG}. A more general form of SHTC equations involving an extra mass flux term is shown (including an analogical derivation) in \ref{sec.rho0}.

Although the distortion was defined as inverse of the deformation gradient, meaning that 
$\partial_i \A{L}{l} = \partial_l \A{L}{i}$, we have actually never used this property. This is the 
crucial point making distortion advantageous, since by including dissipation this condition can be 
violated, i.e.
\begin{equation}\label{torsion}
	\mathfrak{T}^\sL_{ij} = \partial_i \A{L}{j} - \partial_j \A{L}{i}\neq 0 
	\qquad \mbox{or}\qquad 
	\nabla\times \AA \neq 0.
\end{equation}
Tensor $\mathfrak{T}^\sL_{ij}$ is called torsion  tensor 
\cite{PRD-Torsion2019}, and it expresses incompatibility in the deformation 
field. Its physical interpretation depends on the physical context. Usually 
it is interpreted as the dislocation density (or Burgers) tensor 
\cite{Landau7,PRD-Torsion2019} but also can be used to represent the spin of 
the distortion field $ \AA $ which can be associated with the small-scale 
vortex dynamics in turbulent flows as discussed in \cite{PRD-Torsion2019}.
The Lagrangian configuration is then no longer uniquely determined because 
integration of $\AA$ over a closed loop does not necessarily yield zero, which 
is how dislocations are naturally incorporated into the mechanics, e.g. see 
\cite{Yavari2012}. Hence, we have equipped Eulerian coordinates with natural 
state variables which do not dwell on the existence of a continuous mapping 
connecting the reference Lagrangian and actual configurations and allow to 
include formation 
and propagation of defects in continuous medium.

\begin{remark}
It is also important to note that Poisson bracket \eqref{eq.PB.A} satisfies 
Jacobi identity \eqref{eq.Jacobi} unconditionally, see \cite{SHTC-GENERIC}. 
Bracket \eqref{eq.PB.A} gives evolution equations compatible with the evolution equation for $\FF$, Eq. \eqref{eq.evo.Eu.F}.
If the terms proportional to $\nabla\times\AA$ were dropped, the compatibility would be lost, as well as Galilean invariance and unconditional validity of Jacobi identity. 
It is important to keep the terms although they are typically zero in elastic evolution.
\end{remark}

\begin{remark}
Note also that equations \eqref{eq.evo.A} (or \eqref{eq.evo.Eu}) do not represent a system of conservation laws. While the equation for density, momentum density and entropy density can be written as conservation laws, the equation for distortion can not (without neglecting the terms proportional to $\nabla\times\AA$).  This is similar to the evolution equation for $\BB$, Eq. \eqref{eq.B.evo}, which does not represent any conservation law either. In general, there are seven conserved quantities (momentum, angular momentum and energy) given by symmetries of the space-time, \cite{Landau1}. Conservation of mass is added to these conservation laws in classical physics and conservation of entropy is the limiting case of non-dissipative dynamics. Apart from these quantities there is no other universal conserved quantity in classical physics (although various conserved quantities can be derived from those universally conserved, and some conserved quantities appear in special cases \cite{Holm1983}). Therefore, taking finite elasticity into account (meaning state variables $\FF$, $\BB$ or $\AA$ in the Eulerian frame) one can not in general expect to get a system of conservation laws.
\end{remark}

\subsection{Jacobi identity}\label{sec.Jacobi}
Jacobi identity \eqref{eq.Jacobi} is an inherent property of Poisson brackets, 
explicit verification of which is usually a formidable task. This difficulty 
was overcome by program \cite{kroeger2010} checking the identity in an 
automatized way. What is the reason for such interest in Jacobi identity? We 
address this question in the following sub-sections.

\subsubsection{Self-consistency of Hamiltonian dynamics}
Hamiltonian evolution of state variables $\qq$ can be expressed by Eq. 
\eqref{eq.qL}, and from the geometric point of view it can be seen as motion in 
the state space where $\qq$ belongs. The curves $\qq(t)$ are integral curves of 
the \textit{Hamiltonian vector field} $ \XXXX_E $ the components of which 
represent the 
right hand side of Eq. \eqref{eq.qL}, 
\begin{equation}\label{eq.Ham.field}
	\XXXX_E \stackrel{\mathrm{def}}{=} L^{\alpha\beta}E_{q^\beta} \bm{\pd}_\alpha, \quad \XXX^\alpha_E = 
	L^{\alpha\beta}E_{q^\beta},
	\quad 
	\bm{\pd}_\alpha \stackrel{\mathrm{def}}{=} \frac{\partial}{\partial q^\alpha}, 
\end{equation}
or
\begin{equation}\label{eq.Ham.sys}
    \dot{q}^\alpha = \{q^\alpha,E\} = L^{\alpha\beta}E_{q^\beta} = (\LL \cdot dE) (q^\alpha) = \XXXX_E(q^\alpha) = \XXX^\alpha_E.
\end{equation}
Hamiltonian evolution can be seen as motion along the Hamiltonian vector field 
generated by energy $E$.

Having the Hamiltonian vector field, let us ask the question whether the structure of Eq. \eqref{eq.Ham.sys} is kept during the evolution.
Taking arbitrary functionals $\Ffunc$, $\Gfunc$ and $E$, we have
\begin{eqnarray}
\{\{\Ffunc,\Gfunc\},E\}&=&\Lie_{\XXXX_E}(d\Ffunc \cdot \LL \cdot d\Gfunc)  
\nonumber\\
&=&\Lie_{\XXXX_E} (d\Ffunc) \cdot \LL \cdot d\Gfunc
+d\Ffunc \cdot \Lie_{\XXXX_E} (\LL) \cdot d\Gfunc
+d\Ffunc \cdot \LL \cdot \Lie_{\XXXX_E} (d\Gfunc) \nonumber\\
&=&
d \Lie_{\XXXX_E} \Ffunc \cdot \LL \cdot d\Gfunc
+d\Ffunc \cdot \Lie_{\XXXX_E} (\LL) \cdot d\Gfunc
+d\Ffunc \cdot \LL \cdot d \Lie_{\XXXX_E} \Gfunc \nonumber\\
&=&\{\{\Ffunc,E\},\Gfunc\} +d\Ffunc \cdot \Lie_{\XXXX_E} (\LL) \cdot d\Gfunc + 
\{\Ffunc,\{\Gfunc,E\}\},
\end{eqnarray}
where $\Lie_{\XXXX_E}$ is the Lie derivative with respect to the Hamiltonian vector field, see e.g. \cite{Fecko}, which commutes with differential $d$. 
Using Jacobi identity, Eq. \eqref{eq.Jacobi}, we obtain that 
\begin{equation}
d\Ffunc \cdot \Lie_{\XXXX_E} (\LL) \cdot d\Gfunc = 0 \quad\forall \Ffunc, \Gfunc,
\mbox{which means that}\quad 
\Lie_{\XXXX_E}\LL = 0.
\end{equation}
We have thus proved the following proposition, c.f. \cite{Marle2014}, 
\begin{prop}\label{prop.Lie.bivector}
Lie derivative of Poisson bivector along the Hamiltonian vector field, given by Eq. \eqref{eq.Ham.field}, is zero,
\begin{equation}\label{eq.Jac.Lie}
	\Lie_{\XXXX_E}\LL = 0,
\end{equation}
see e.g. \cite{Marle2014}.
\end{prop}
This tells that the Poisson bivector $\LL$ does not change along the evolution of the system. Jacobi identity can be seen as a condition of self-consistency of the reversible Hamiltonian evolution. As a corollary, the canonical Poisson bivector, which is a constant tensor field and thus has zero Lie derivative, automatically satisfies Jacobi identity.

\subsubsection{Criterion when constructing Poisson brackets}
Jacobi identity is also useful as a decisive criterion when choosing between 
several possible forms of a Poisson bracket. For instance, in 
\cite{Markus2009}, it led to the identification of coupling between the 
mechanics of the Eulerian deformation gradient $\FF(\xx)$ and fluid mechanics. 
Similarly, in \cite{SHTC-GENERIC}  bracket \eqref{eq.PB.A} is derived by 
reduction of a simpler bracket for fluid mechanics with labels (distortion 
being spatial gradient of labels). By adding terms to the bracket that are zero 
for compatible distortion matrices ($\nabla\times\AA=0$), Jacobi identity 
becomes valid unconditionally (even with incompatible distortion), and bracket 
\eqref{eq.PB.A} is recovered.

In \cite{Miroslav-Grad}, the Poisson bracket for the infinite Grad hierarchy in 
kinetic theory was formulated. Projection for instance to the first ten moments 
(fluid mechanics and the matrix of second moments) does not end up in a closed 
form. Jacobi identity can be seen as a closure criterion so that the resulting 
evolution equations become frame invariant. 

\subsection{Gauge invariance, symmetries and conserved quantities}\label{sec.gauge}

We shall now make a few observations regarding conserved quantities, symmetries and transformation invariants in Hamiltonian systems. These links and properties follow again from the structure of Poisson bracket and allow stronger statements about symmetries and conservation laws than in Lagrangian systems observed by Noether. The results can be obtained as corollaries of \cite{olver2000applications}, but here we propose an alternative viewpoint that allows a significantly shorter exposition and reveals different relationships and insight. 

We approach this problem from two perspectives. The first one, Sec. \ref{sec.Noether}, \ref{sec.symmetry} and \ref{sec.conserved}, is rather 
intuitive, easily understandable and invoking the properties of Hamiltonian 
systems but rather formal.
Subsequently, in Sec. \ref{sec.rigorous}, we built upon rigorous results from \cite{olver2000applications}.

\subsubsection{Noether (and inverse) theorem for Poisson systems}\label{sec.Noether}
Assume now that there is a conserved quantity $\Gfunc(\qq)$ of the Hamiltonian 
system \eqref{eq.Ham.sys}, i.e. $\{\Gfunc,E\}=0$. This property can be 
rewritten as 
\begin{equation}\label{eq.symmetry}
    0 = \{\Gfunc,E\}= \Gfunc_{q^\alpha} L^{\alpha\beta} E_{q^\beta} = 
    \XXXX_E(\Gfunc)=\Lie_{\XXXX_E}\Gfunc,
\end{equation}
which means that the action of the vector field on functional $\Gfunc$ 
(conserved quantity) is zero. 
From antisymmetry of the Poisson bracket we also have
\begin{equation}\label{eq.sym.back}
    0 = \{E,\Gfunc\}= E_{q^\alpha} L^{\alpha\beta} \Gfunc_{q^\beta} = 
    \XXXX_\Gfunc(E)=\Lie_{\XXXX_\Gfunc}E,
\end{equation}
which means that field $\XXXX_\Gfunc$ represents a symmetry of the Hamiltonian in 
the following sense.
\begin{definition}[Symmetry of Hamiltonian]
	A vector field $\XXXX$ is a symmetry of Hamiltonian $E$ when its action on 
	the Hamiltonian is zero, $\Lie_{\XXXX}E=\XXXX(E)=0$.
\end{definition}
Looking at Eqs. \eqref{eq.symmetry} and \eqref{eq.sym.back}, we obtain the Hamiltonian version of famous Noether theorem.
\begin{theorem}[Emmy Noether, and its inverse for Poisson systems]
Any conserved quantity $\Gfunc(\qq)$ of Hamiltonian system \eqref{eq.Ham.sys} 
generates a Hamiltonian vector field $\XXXX_\Gfunc$, which is a symmetry of the 
Hamiltonian, $\Lie_{\XXXX_\Gfunc} E = 0$. Conversely, if a Hamiltonian vector 
field $\XXXX_\Gfunc$ is a symmetry of the Hamiltonian, then generator $\Gfunc$ of 
the field 
is conserved. See also \cite{Butterfield}.
\end{theorem}

\subsubsection{Symmetry of a Hamiltonian dynamical system}\label{sec.symmetry}
Let us now focus on infinitesimal symmetries of Hamiltonian dynamical systems \eqref{eq.Ham.sys}. Note that the calculations are formal and well substantiated only for finite dimensional systems. However, we shall proceed in this formal treatment as it elucidates the geometric content.

A Hamiltonian vector field $\XXXX_\Gfunc$ defines an infinitesimal transformation 
of state variables 
\begin{equation}\label{eq.barq}
    \bar{q}^\alpha = q^\alpha + \eps \XXXX_\Gfunc(q^\alpha) = q^\alpha + \eps \Lie_{\XXXX_\Gfunc} q^\alpha = 
    q^\alpha 
    + \eps \{q^\alpha, \Gfunc\}.
\end{equation}
This formula actually represents infinitesimal transformations in a broader sense, see Sec. \ref{sec.rigorous} for a rigorous treatment.

For Lie derivatives it holds that, see e.g. \cite{Fecko},
\begin{equation}
    \Lie_{[\XXXX,\YYYY]} = \Lie_{\XXXX} \Lie_{\YYYY}-\Lie_{\YYYY} \Lie_{\XXXX}
\end{equation}
for any vector fields $\XXXX$ and $\YYYY$. Therefore, assuming that a vector field $\XXXX$ commutes with $\XXXX_E$, $[\XXXX,\XXXX_E]=0$, it follows that 
\begin{equation}\label{eq.LieLie}
\Lie_{\XXXX} \Lie_{\XXXX_E} = \Lie_{\XXXX_E} \Lie_{\XXXX}.
\end{equation}

Considering Hamiltonian system \eqref{eq.Ham.sys}, solution after an infinitesimal time $dt$ is equal to 
\begin{equation}
    q^\alpha(t+ dt) = q^\alpha(t) + dt \Lie_{\XXXX_E} q^\alpha(t) 
\end{equation}
with correction terms of order $(dt)^2$. A vector field $\XXXX$ generates infinitesimal transformation \eqref{eq.barq}. Assuming that it commutes with $\XXXX_E$, i.e. identity \eqref{eq.LieLie} holds,  the transformed variables at time $t+dt$ are
\begin{eqnarray}
   \bar{q}^\alpha(t+dt) &=& q^\alpha(t+dt) + \eps \Lie_{\XXXX} q^\alpha(t+dt)\nonumber\\
   &=& q^\alpha(t) + dt \Lie_{\XXXX_E} q^\alpha(t) +\eps \Lie_{\XXXX} q^\alpha(t) + \eps \Lie_{\XXXX} (dt \Lie_{\XXXX_E} q^\alpha(t))\nonumber\\
   &=&q^\alpha(t) +\eps \Lie_{\XXXX} q^\alpha(t) +dt \Lie_{\XXXX_E}\left(q^\alpha(t) + \eps \Lie_{\XXXX} q^\alpha(t)\right)\nonumber\\
   &=& \bar{q}^\alpha(t) + dt \Lie_{\XXXX_E} \bar{q}^\alpha(t).
\end{eqnarray}
Therefore, the transformed quantities \eqref{eq.barq} obey the same evolution equations as the quantities before transformation, and thus they have the same set of solutions. Such a transformation is called a symmetry of the dynamical system. Hence we have shown the following proposition. 
\begin{prop}[Symmetries of Hamiltonian system] 
A symmetry of the dynamical system is any vector field $\XXXX$ that commutes with the Hamiltonian vector field, 
\begin{equation}\label{eq.def.sym.sys}
    0 = [\XXXX_E,\XXXX] = \XXXX_E \XXXX - \XXXX \XXXX_E = \Lie_{\XXXX_E} \XXXX = -\Lie_{\XXXX} \XXXX_E.
\end{equation}
The symmetry induces an infinitesimal transformation \eqref{eq.barq}, 
 after which the Hamiltonian system has the same set of solutions (in short 
 time) as the original system, see also \cite{olver2000applications}.
\end{prop}

\subsubsection{Symmetries of Hamiltonian systems and conserved quantities}\label{sec.conserved}
Finally, let us now turn to the question whether a conserved quantity also represents a symmetry of the Hamiltonian dynamical system. Using the definition of the symmetry, Eq. \eqref{eq.def.sym.sys}, we have
\begin{eqnarray}\label{eq.EG.com}
- \Lie_{\XXXX_\Gfunc} \XXXX_E &=& \Lie_{\XXXX_E}\XXXX_\Gfunc = \Lie_{\XXXX_E} (\LL\cdot 
d\Gfunc) = 
\Lie_{\XXXX_E}\LL \cdot d\Gfunc +\LL\cdot\Lie_{\XXXX_E} d\Gfunc\nonumber\\
	&=& \Lie_{\XXXX_E} \LL  \cdot d\Gfunc + \LL \cdot d \underbrace{\Lie_{\XXXX_E} 
	\Gfunc}_{=0} = \underbrace{\Lie_{\XXXX_E} \LL}_{=0}  \cdot d\Gfunc=0,
\end{eqnarray}
where we used that the Lie derivative commutes with differential, $\Lie d = d 
\Lie$, see e.g. \cite{Fecko}, and where Jacobi identity was used, see 
Proposition\,\ref{prop.Lie.bivector}. From this observation it follows that if 
a functional 
$\Gfunc$ 
is a conserved quantity, $\dot{\Gfunc}=0$, of a Hamiltonian dynamical system 
\eqref{eq.Ham.sys}, the Hamiltonian vector field generated by $\Gfunc$ is a 
symmetry 
of the Hamiltonian system. Using this identity in Eq. \eqref{eq.EG.com}, we 
obtain the following theorem.
\begin{theorem}\label{theorem.cons.sym}
Assume that a functional $\Gfunc$ is a conserved quantity of a Hamiltonian 
dynamical system \eqref{eq.Ham.sys}, see Eq. \eqref{eq.symmetry}. Then the 
Hamiltonian vector field generated by $\Gfunc$ is a symmetry of the Hamiltonian 
system in the sense of Eq. \eqref{eq.def.sym.sys}, cf. \cite{olver2000applications}.
\end{theorem}

Hence we not only know 
that a conservation law generates a 
symmetry, but we know the relation explicitly, i.e. a direct relation between a 
conserved quantity and an invariance in the system. This result can be extended 
to a general case, but certain technical extensions of the concepts here have 
to be carried out \cite{olver2000applications}, e.g. the relation 
\eqref{eq.barq} is no longer a transformation (strictly speaking), but can be 
shown to be a prolongation of a Hamiltonian vector field, see \ref{sec.rigorous} for more details.


For a Hamiltonian system, it is possible to prove invariance of the evolution equations with 
respect to an infinitesimal transformation by showing the conservation of the transformation 
generator. For instance, Galilean invariance can be shown relatively easily by proving 
conservation of Galilean booster, and the method is not restricted to evolution equations in the 
form of conservation laws, see \ref{sec.rigorous} for concrete examples.

\subsection{Hyperbolicity}\label{sec.hype}
Hyperbolicity is an essential feature of many systems in continuum thermodynamics 
\cite{Muller-Ruggeri,Fischer1972,Kremer-14,Sbierski2016}, since it provides well-posedness of the Cauchy problem 
locally in time and causality, see Sec.\,\ref{sec.intro}. However, it is not easy to check 
hyperbolicity when the equations 
are not in the form of system of conservation laws admitting an extra conservation law (typically 
energy conservation), see e.g. \cite{SHTC-GENERIC} and the Godunov-Boillat theorem, first 
proposed by Godunov  \cite{Godunov-interesting}, and generalized in \cite{Ruggeri1989}, 
\cite{Boillat1974}, \cite{Ruggeri-Strumia} and  \cite{FriedLax}. Therefore, our motivation for this 
section  is to shed some light on interconnections between the symmetric hyperbolicity (particular 
case of hyperbolicity) and Hamiltonian formulation of a system of conservation laws. Note that in 
this section no summation over repeated indexes is meant.

\subsubsection{Godunov-Boillat theorem}
The usual way to show symmetric hyperbolicity of a system of quasilinear first order equations is the Godunov-Boillat theorem, see e.g. \cite{Godunov-interesting,Boillat1974,Romenskii-Maxwell,SHTC-GENERIC}. The theorem is based on the passage to a dual formulation by means of Legendre transformation 
\begin{equation}\label{eq.qp}
q^\dagger_\alpha = \frac{\partial e}{\partial q^\alpha}, \qquad q^\dagger_\alpha = \frac{\partial L}{\partial q^\dagger_\alpha}
\qquad\mbox{and}\qquad L = -e + q^\alpha q^\dagger_\alpha,
\end{equation}
and it works for systems of overdetermined conservation laws (automatically implying an extra 
conservation law, e.g. energy conservation). Note that $L$ has the meaning of pressure, cf. Eq. \eqref{eq.p}, and it is the complete Legendre transformation of energy density. For non-conservative systems, as for instance the SHTC 
equations in Section \ref{sec.SHTC}, the theorem can be applied either by restriction to the 
compatible systems ($\curl \AA = 
0$) or by extension promoting the incompatibility to an extra state variables (Burgers tensor), see 
\cite{Romenskii-Maxwell,SHTC-GENERIC}. The way based on the Hamiltonian structure of the equations, Eqs. 
\eqref{sys.hydro3}, can be seen as an alternative (or extension) to the Godunov-Boillat theorem (the theorem is recalled in \ref{sec.GBproof}).

\subsubsection{Riemannian approach}
It is sometimes possible to infer hyperbolicity just from the Hamiltonian character of 
the equations. Let us first recall some results by Dubrovin, Novikov \cite{Novikov} and Tsar{\" e}v 
\cite{Tsarev}. 
A Poisson bracket in 1D is of \textit{hydrodynamic type} if the corresponding Poisson bivector has 
the form
\begin{equation}\label{eqn.bivector.hydro}
	L^{\alpha\beta} = \{q^\alpha(x),q^\beta(y)\} = g^{\alpha\beta}(\qq(x)) \partial_x \delta(x-y) + b^{\alpha\beta}_\gamma (\qq(x)) \partial_x q^\gamma \delta(x-y).
\end{equation}
For further discussion see \cite{Dubrovin-Novikov}.
Energy is of hydrodynamic type if it is an integral of a function of the state variables, 
\begin{equation}
	E = \int \diff x e(\qq(x)).
\end{equation}
The evolution equations generated by a hydrodynamic-type Poisson bracket and hydrodynamic-type energy are quasilinear partial differential equations of first order
\begin{equation}\label{eqn.hydro.type}
	\partial_t q^\alpha = \vab{\alpha}{\beta}(\qq(x)) \partial_x q^\beta
\end{equation}
with
\begin{equation}
	\vab{\delta}{\gamma} = g^{\delta\beta} \frac{\partial^2 e}{\partial q^\beta \partial q^\gamma} + b^{\delta\beta}_\gamma \frac{\partial e}{\partial q^\beta}
\end{equation}
as follows by direct calculation.
The Poisson brackets discussed in this paper are all of hydrodynamic type.


It was shown in \cite{Novikov} that
\begin{enumerate}
\item Under local changes $\QQ = \QQ(\qq)$ of the vector of state variables, the coefficient $g^{\alpha\beta}$ in \eqref{eqn.bivector.hydro} is transformed as a tensor with upper indices; if $\det g^{\alpha\beta} \neq 0$, then the expression, cf. Eq. \eqref{eqn.bivector.hydro},
		$b^{\alpha\beta}_\gamma = -g^{\alpha\delta}\Gamma^\beta_{\delta\gamma}$ 
		is transformed so that $\Gamma^\beta_{\delta\gamma}$ is the Christoffel symbol of an affine connection.
\item  For the Poisson bivector to be antisymmetric it is necessary and sufficient that the
	tensor $g^{\alpha\beta}$ be \textit{symmetric} (i.e. it defines a pseudo-Riemannian metric, if $\det g^{\alpha\beta} \neq 0$) and
		the connection $\Gamma^\alpha_{\beta\gamma}$ is metric compatible: $\nabla_\gamma g^{\alpha\beta} = 0\, \forall \alpha,\beta,\gamma$,
		where $\nabla_\gamma$ is the associated 
		covariant derivative in the space of state variables.
	\item  For the bracket to satisfy Jacobi identity it is necessary and sufficient (in the case $\det
	g^{\alpha\beta} \neq 0 $) that the connection $\Gamma^{\beta}_{\delta\gamma}$ be \textit{torsion-} and \textit{curvature-free}.
\end{enumerate}
Assuming the non-degenerate case, the covariant Hessian of energy can be rewritten as 
\begin{equation}
    \nabla_\delta \nabla_\gamma e = \frac{\partial}{\partial q^\delta}\frac{\partial e}{\partial q^\gamma} - \Gamma^\beta_{\delta\gamma}\frac{\partial e}{\partial q^\beta},
\end{equation}
from which it follows that the matrix $\vab{\delta}{\gamma}$ can be rewritten as
\begin{equation}
	\vab{\gamma}{\delta} = g^{\delta\beta}\nabla_\beta \nabla_\delta e.
\end{equation}
Furthermore, let us assume that the energy $e(\qq)$ is a proper scalar (i.e. $e(\qq) = e(\QQ(\qq))$). Then, due to that there is neither torsion nor curvature, the covariant derivatives commute and the matrix
$e_{\delta\gamma}\stackrel{\mathrm{def}}{=} \nabla_\delta \nabla_\gamma e$ is symmetric. The evolution equations \eqref{eqn.hydro.type} then become
\begin{equation}\label{sys.hydro2}
	\partial_t q^\alpha = g^{\alpha\beta} e_{\beta\gamma}\partial_x q^\gamma,
\end{equation}
which can be symmetrized by multiplying it by the covariant Hessian of energy $e_{\delta\alpha}$. One obtains
\begin{equation}\label{sys.hydro3}
	e_{\delta\alpha}\partial_t q^\alpha = e_{\delta\alpha} g^{\alpha\beta} e_{\beta\gamma}\partial_x q^\gamma.
\end{equation}
If the energy is convex, then it follows that when taking a curve in the space of state variables, 
second derivative of energy with respect to a parameter parametrizing the curve is positive. This 
holds true in particular for geodesic curves. Therefore, energy is also geodesic convex 
\cite{Vishnoi} and its covariant Hessian is positive definite, i.e. symmetric positive definite. 
Equations \eqref{sys.hydro2} are thus equivalent (at least regarding their strong solutions) to the 
original system \eqref{sys.hydro2}, the matrix in front of the time-derivative is symmetric 
positive-definite, and the matrices in front of the spatial derivatives are symmetric. Equations 
\eqref{sys.hydro3} thus form a system of quasilinear first-order \textit{symmetric hyperbolic} 
partial differential equations \cite{Friedrichs1958,God1961}. These results can be summarized as 
the following theorem:
\begin{theorem}
Consider a one-dimensional Hamiltonian system of hydrodynamic type with non-degenerate metric 
\eqref{eqn.bivector.hydro}. Assuming that the energy of the system is of hydrodynamic type, 
convex and a proper scalar, it follows that the evolution equations can be regarded as a 
first-order quasilinear symmetric hyperbolic 
PDE system.
\end{theorem}

Let us give a few examples. Isentropic fluids in one-dimension are described by the mass density 
and 
momentum density 
$(\rho,m)$ (so $m$ is a one-dimensional covector field) and thus, the first two terms in Poisson 
bracket \eqref{eq.PB.FM}, have the metric 
\begin{equation}
    \mathbf{g} = \begin{pmatrix}0 & -\rho \\ -\rho & -2m\end{pmatrix}.
\end{equation}
This metric is non-degenerate and symmetric, but also indefinite. 

To include entropy, one has to add not only the entropy field, but also the field of conjugate 
entropy flux $\ww$ (see \cite{SHTC-GENERIC}), since otherwise the metric would be degenerate 
\cite{Dubrovin-Novikov}. The Poisson bracket, including $s$ and $\ww$ fields, is
\begin{eqnarray}\label{eq.PB.sw}
    \{\Ffunc,\Gfunc\}&=&\{\Ffunc,\Gfunc\}^{(FM)} + \int\diff\xx (\nabla \Ffunc_s 
    \cdot \Gfunc_\ww - \nabla \Gfunc_s \cdot \Ffunc_\ww)\nonumber\\
&&+\int\dx w_k \left(\Ffunc_{w_j} \partial_j \Gfunc_{m_k}-\Gfunc_{w_j} \partial_j \Ffunc_{m_k}\right)\nonumber\\
&&-\int\dx \partial_k w_j \left(\Ffunc_{w_j} \Gfunc_{m_k}-\Gfunc_{w_j} \Ffunc_{m_k}\right),
\end{eqnarray}
which is analogical to the bracket involving $\rho$ and $\ww$ in \ref{sec.rho0}, taking $s$ instead of $\rho$.
Bracket \eqref{eq.PB.sw} leads to the metric (again in 1D, i.e. only first components $ w $ and $ m $ of $ \ww $ and $ 
\mm $ are considered)
\begin{equation}
    \mathbf{g} = 
    \begin{pmatrix}
    0 & -\rho & 0 & 0\\ 
    -\rho & -2m & -s & -w\\
    0 & -s & 0 & -1\\
    0 & -w &  -1 & 0
    \end{pmatrix},
\end{equation}
which is again symmetric, indefinite and non-degenerate. 

Similarly, Poisson bracket \eqref{eq.PB.A} restricted to functionals dependent only on $(\mm,\AA)$ 
has non-degenerate metric. Entropy can be added by including also the $\ww$ field as above. Density 
can be added either by its relation with $\det (\AA)$ or by adding both density and a conjugate 
velocity-like field (similar to the $\ww$ field) coupled to it, see \ref{sec.rho0} or \cite{Peshkov-Grmela}. 

Let us now bring the two approaches to proving hyperbolicity (Riemannian and Godunov-Boillat) closer to each other.
If $e_{\beta\gamma}$ were the usual derivatives (not covariant), i.e. $e_{\beta\gamma}=\pd_\beta\pd_\gamma e$, the situation 
would be simple. Taking again $p_\alpha = e_{q^\alpha}$ and $L= -e + p_\alpha q^\alpha $, then $p_\alpha = 
L_{p_\alpha}$ and the system of equations becomes 
\begin{equation}
    \pd_t L_{p_\alpha} = g^{\alpha\beta}e_{\beta\gamma}\pd_x L_{p_\gamma},
\end{equation}
or
\begin{equation}
    L^{\alpha\delta}\pd_t p_\delta = g^{\alpha\beta}e_{\beta\gamma}L^{\gamma\delta}\pd_x p_\delta, 
    \qquad\mbox{where}\quad L^{\alpha\delta} = L_{p_\alpha p_\delta}.
\end{equation}
Because $e_{\beta\gamma}L^{\gamma\delta} = \delta^\delta_{\ \gamma}$, we would have a symmetric hyperbolic system
\begin{equation}
    L^{\alpha\delta}\pd_t p_\delta = g^{\alpha\beta}\pd_x p_\beta.
\end{equation}
However, the Hessian $e_{\beta\gamma}$ is made of covariant derivatives, not partial. By this remark, we 
would like to 
draw attention to this similarity between the two approaches (Godunov-Boillat and Riemannian). 

\begin{remark}
Moreover, Theorem 1 from \cite{Dubrovin-Novikov} tells that once the metric is non-degenerate, then there exists such a local change of variables that the metric becomes pseudo-Euclidean and $b^{ij}_k=0$. Then the covariant derivatives become partial derivatives and the symmetrization by the Godunov-Boillat theorem becomes equivalent to the symmetrization by the Riemannian approach.
\end{remark}
%

In summary, there is a new link between Hamiltonian systems of hydrodynamic type and their 
symmetric hyperbolicity. If the metric constructed from the Poisson bivector is non-degenerate and 
if the energy is convex, then the resulting evolution equations are symmetric hyperbolic. This 
provides an alternative to the Godunov-Boillat theorem, especially in the case of non-conservative 
equations.

\subsection{Clebsch variables}\label{sec.Clebsch}
Variational principles for fluid mechanics have been of great importance in physics. Clebsch 
\cite{Clebsch} found the canonical variables providing Hamiltonian structure to fluid mechanics, 
Seliger, Whitham \cite{Seliger-Whitham} and Lin \cite{Lin} equipped fluid mechanics with labels to 
gain the variational structure, \cite{Bedeaux-Clebsch}. See \cite{Cendra-Marsden} and 
\cite{Cotter2009} for a clearer and more geometric explanation of the results. For instance in 
\cite{PKG}, the Clebsch variables were written as $\rho(\xx)$, $\rho^*(\xx)$, $\lambda(\xx)$, 
$\lambda^*(\xx)$, $s(\xx)$ and $s^*(\xx)$ equipped with the canonical Poisson bracket for fields,
\begin{multline}
    \{\Ffunc,\Gfunc\}^{(Clebsch)} = \int\dx \left(\frac{\delta \Ffunc}{\delta 
    \rho}\frac{\delta \Gfunc}{\delta \rho^*}-\frac{\delta \Gfunc}{\delta 
    \rho}\frac{\delta \Ffunc}{\delta \rho^*}\right)\\
    + \int\dx \left(\frac{\delta \Ffunc}{\delta \lambda}\frac{\delta 
    \Gfunc}{\delta 
    \lambda^*}-\frac{\delta \Gfunc}{\delta \lambda}\frac{\delta \Ffunc}{\delta 
    \lambda^*}\right)\\
    + \int\dx \left(\frac{\delta \Ffunc}{\delta s}\frac{\delta \Gfunc}{\delta 
    s^*}-\frac{\delta \Gfunc}{\delta s}\frac{\delta \Ffunc}{\delta s}\right)
\end{multline}
for all $\Ffunc$ and $\Gfunc$ smooth enough functionals of the Clebsch variables. The evolution 
equations implied by this canonical bracket are
\begin{equation}
    \partial_t \rho = \frac{\delta E}{\delta \rho^*}, \quad \partial_t \rho^* = - \frac{\delta 
    E}{\delta \rho}, \quad \mathrm{etc.},
\end{equation}
and they can be seen as a consequence of the principle of least action (with variations vanishing at boundaries)
\begin{equation}
    \delta \int_{t_1}^{t_2} \int\dx L(\rho,\dot{\rho}, \lambda, \dot{\lambda}, s, \dot{s}) = 0,
\end{equation}
where $L$ is the Lagrangian density related to energy by the Legendre transformation.

Fluid mechanics is then obtained by the reduction
\begin{subequations}
\begin{eqnarray}
    \rho &=& \rho,\\
    m_i &=& \rho\partial_i \rho^* + \lambda \partial_i \lambda^* + s \partial_i s^*,\\
    s &=& s,
\end{eqnarray}
\end{subequations}
under which the canonical Clebsch Poisson bracket turns to the Poisson bracket for fluid mechanics 
\eqref{eq.PB.FM}.

One can, however, project the Clebsch variables not only to the fluid mechanics, but also to keep the $\lambda$ field, which can be seen as volume density of labels. Indeed, the reduction then leads to Poisson bracket
\begin{equation}
    \{\Ffunc,\Gfunc\}^{(FM)} + \int\dx \lambda (\nabla \Ffunc_\lambda \cdot \Gfunc_\mm-\nabla 
    \Gfunc_\lambda \cdot 
    \Ffunc_\mm),
\end{equation}
which implies the evolution equation for $\lambda$
\begin{equation}
    \partial_t \lambda = -\partial_j (\lambda E_{m_j}).
\end{equation}
Subsequent change of variables to $X = \lambda/\rho$ then yields 
\begin{equation}
    \partial_t X = -E_{m_i} \partial_i X,
\end{equation}
which is a simple advection of function $X$ (a marker or a label) by the fluid. Starting with the 
three 
fields $\lambda^\sI$, where $\sI$ is a Lagrangian index, the resulting Poisson bracket (called Lin 
bracket) is 
\begin{equation}
\{\Ffunc,\Gfunc\}^{(Lin)} = \{\Ffunc,\Gfunc\}^{(FM)} + \int\dx \partial_j X^\sI (\Ffunc_{m_j} 
\Gfunc_{X^\sI}-\Gfunc_{m_j} \Ffunc_{X^\sI}),
\end{equation}
see \cite{Grmela-PLA,PKG} for more details.
The Lin Poisson bracket yields evolution equations for fluid mechanics equipped with labels
\begin{subequations}\label{eq.evo.Lin}
\begin{eqnarray}
	\partial_t \rho &=& -\partial_i(\rho E_{m_i})\\
	\partial_t m_i &=& -\rho\partial_i E_\rho - m_j \partial_i E_{m_j} -s \partial_i E_s 
	-\partial_i X^\sI \\
	\partial_t s &=& -\partial_i (s E_{m_i})\\
	\partial_t X^\sI &=& -E_{m_i}\partial_i X^\sI.
\end{eqnarray}
\end{subequations}

Finally, the field of labels $X^\sI$ can be projected to the distortion matrix through
\begin{equation}\label{eq.A.a}
	\A{I}{i} = \frac{\partial X^\sI}{\partial x^i},
\end{equation}
and the consequent reduction of the Lin Poisson bracket leads to the Poisson bracket for 
distortion matrix \eqref{eq.PB.Eu}. Note however, that the terms proportional to $\nabla\times \AA$ 
do not appear in the result automatically. They are zero provided the construction \eqref{eq.A.a} 
is smooth enough, which is why they do not appear. However, the Jacobi identity is then fulfilled 
provided that $\nabla\times \AA=0$, and the terms in \eqref{eq.PB.Eu} proportional to $\nabla\times 
\AA$ 
have 
to be added to make the Jacobi identity fulfilled unconditionally, see \cite{SHTC-GENERIC}.

In summary, we show that Clebsch variables provide an alternative formulation of variational principle for fluid mechanics and fluid mechanics with distortion.

\subsection{Semidirect product structure}\label{sec.SP}
It is well known that mechanics (i.e. reversible evolution) of fluids is Hamiltonian, see e.g. 
\cite{Arnold,Marsden-Ratiu-Weinstein} or \cite{PKG}. Fluid mechanics, in particular, is a 
realization of Lie-Poisson dynamics, where the Poisson bracket is the Lie-Poisson bracket on a Lie 
algebra dual. Another examples of Lie-Poisson dynamics are rigid body rotation or kinetic theory. 
In \cite{Marsden-Ratiu-Weinstein} it is explained how to construct new Hamiltonian dynamics by 
letting one Hamiltonian dynamics be advected by another. Having a Lie algebra dual $\LA^*$ (for 
instance fluid mechanics), an another Lie algebra dual or cotangent bundle is advected by $\LA^*$ 
by the construction of semidirect product. 

One can even think of mutual action of the two Hamiltonian dynamics, which leads to the structure of matched pairs \cite{esen2016hamiltonian}, \cite{elmag}. For the purpose of this paper, however, we restrict the discussion only to one-sided action of one Hamiltonian system to another, i.e. to the semidirect product. A general formula for the Poisson bracket of semidirect product of a Lie algebra dual $\LA^*$ and cotangent bundle $T^*M = V\times V^*$ was presented for instance in \cite{elmag, Ogul-matched2,Affro-elchem},
\begin{eqnarray}\label{eq.PB.SP.gen}
	\{\Ffunc,\Gfunc\}^{(\LA*\ltimes T^*M)} &=& \{\Ffunc,\Gfunc\}^{(\LA^*)} + 
	\{\Ffunc,\Gfunc\}^{(T^*M)}\nonumber\\
	&&+\left\langle  \Ffunc_\AA|\Gfunc_\mm \vartriangleright \AA\right\rangle
	-\left\langle  \Gfunc_\AA|\Ffunc_\mm \vartriangleright \AA\right\rangle\nonumber\\
	&&+ \left\langle \DD| \Ffunc_\mm\vartriangleright  \Gfunc_\DD\right\rangle
	- \left\langle \DD| \Gfunc_\mm\vartriangleright  \Ffunc_\DD\right\rangle
\end{eqnarray}
where $\AA\in V$ is a covector field, $\AA = A_i dx^i$ and $\DD\in V^*$ is a vector field $\DD = 
D^i \partial_i$, $\{\Ffunc,\Gfunc\}^{(\LA^*)}$ is the Lie-Poisson bracket on the Lie algebra dual, 
$\{\Ffunc,\Gfunc\}^{(T^*M)}$ is the canonical Poisson bracket on the cotangent bundle, 
$\langle\bullet|\bullet\rangle$ is a scalar product (usually $L^2$, i.e. integration over the 
domain), $\mm\in\LA^*$ is the momentum density (element 
of the Lie algebra dual) and $\vartriangleright$ is the action of $\LA^*$ on $T^*M$,  minus the Lie 
derivative $- \Lie$. Poisson bracket \eqref{eq.PB.SP.gen} can be thus rewritten as
\begin{eqnarray}
	\{\Ffunc,\Gfunc\}^{(\LA*\ltimes T^*M)} &=& \{\Ffunc,\Gfunc\}^{(\LA^*)} + 
	\{\Ffunc,\Gfunc\}^{(T^*M)}\nonumber\\
	&&-\left\langle  \Ffunc_\AA| \Lie_{\Gfunc_\mm} \AA\right\rangle
	+\left\langle  \Gfunc_\AA| \Lie_{\Ffunc_\mm} \AA\right\rangle\nonumber\\
	&&- \left\langle \DD| \Lie_{\Ffunc_\mm}  \Gfunc_\DD\right\rangle
	+ \left\langle \DD| \Lie_{\Gfunc_\mm} \Ffunc_\DD\right\rangle,
\end{eqnarray}
 with $\{\Ffunc,\Gfunc\}^{(\LA^*)} = \{\Ffunc,\Gfunc\}^{(FM)}$ and 
 \begin{equation}
 \{\Ffunc,\Gfunc\}^{(T^*M)} = \left\langle \Ffunc_\AA, \Gfunc_\DD\right\rangle -\left\langle 
 \Gfunc_\AA, \Ffunc_\DD\right\rangle 
 = \int\dx \left(\frac{\delta \Ffunc}{\delta A_i}\frac{\delta \Gfunc}{\delta 
 D^i}-\frac{\delta \Gfunc}{\delta A_i}\frac{\delta \Ffunc}{\delta D^i}\right).
 \end{equation}
 Lie derivatives of vector and covector fields read (see e.g. \cite{Fecko})
 \begin{subequations}
\begin{eqnarray}
\Lie_\vv \AA &=& \left(v^j \frac{\partial A_i}{\partial x^j} + \frac{\partial v^j}{\partial x^i}A_j\right) dx^i\\
\Lie_\vv \DD &=& \left(v^j \frac{\partial D^i}{\partial x^j} - D^j\frac{\partial v^i}{\partial x^j}\right)\frac{\partial}{\partial x^i},
\end{eqnarray}
\end{subequations}
and the Poisson bracket thus gains the explicit form (noting that $\Ffunc_\DD$ and $\Gfunc_\DD$ are 
covector fields)
\begin{eqnarray}
	\{\Ffunc,\Gfunc\}^{(\LA^*\ltimes T^*M)} &=& \{\Ffunc,\Gfunc\}^{(FM)} 
    + \int\dx \left(\frac{\delta \Ffunc}{\delta A_i}\frac{\delta \Gfunc}{\delta 
    D^i}-\frac{\delta \Gfunc}{\delta A_i}\frac{\delta \Ffunc}{\delta 
    D^i}\right)\\
	&&-\int\dx  \Ffunc_{A_i} (\Gfunc_{m_j} \partial_j A_i + \partial_i \Gfunc_{m_j}A_j)\nonumber\\
	&&+\int\dx  \Gfunc_{A_i} (\Ffunc_{m_j} \partial_j A_i + \partial_i \Ffunc_{m_j}A_j)\nonumber\\
	&&- \int\dx D^i(\Ffunc_{m_j} \partial_j \Gfunc_{D^i} + \partial_i \Ffunc_{m_j} \Gfunc_{D^j} )\nonumber\\
	&&+ \int\dx D^i(\Gfunc_{m_j} \partial_j \Ffunc_{D^i} + \partial_i \Gfunc_{m_j} 
   \Ffunc_{D^j}).\nonumber
\end{eqnarray}
 This Poisson bracket expresses kinematics of a cotangent bundle (a vector field and a covector field) advected by a Lie algebra dual (for instance fluid mechanics).
 
 Let us now equip the covector field $\AA$ with an additional (Lagrangian) index, which is 
 equivalent 
 to letting additional copies of the cotangent bundle be advected by the Lie algebra dual, $A_i 
 \rightarrow \A{I}{i}$. In analogy, the vector field $\DD$ becomes $D^i_{\ I}$, and the Poisson 
 bracket 
 becomes
 \begin{eqnarray}
	\{\Ffunc,\Gfunc\}^{(\LA^*\ltimes T^*M)} &=& \{\Ffunc,\Gfunc\}^{(FM)} 
    + \int\dx \left(\frac{\delta \Ffunc}{\delta \A{I}{i}}\frac{\delta 
    \Gfunc}{\delta 
    D^i_{\ \sI}}-\frac{\delta 
    \Gfunc}{\delta \A{I}{i}}\frac{\delta \Ffunc}{\delta D^i_{\ \sI}}\right)\nonumber\\
	&&-\int\dx  \Ffunc_{\A{I}{i}} (\Gfunc_{m_j} \partial_j \A{I}{i} + \partial_i 
	\Gfunc_{m_j}\A{I}{j})\nonumber\\
	 &&+\int\dx  \Gfunc_{\A{I}{i}} (\Ffunc_{m_j} \partial_j \A{I}{i} + \partial_i 
	\Ffunc_{m_j}\A{I}{j})\nonumber\\
	&&- \int\dx D^i_{\ \sI}(\Ffunc_{m_j} \partial_j \Gfunc_{D^i_{\ \sI}} + \partial_i \Ffunc_{m_j} 
	\Gfunc_{D^j_{\ \sI}}\nonumber\\
	 &&+ \int\dx D^i_{\ \sI}(\Gfunc_{m_j} \partial_j \Ffunc_{D^i_{\ \sI}}) + \partial_i \Gfunc_{m_j} 
   \Ffunc_{D^j_{\ 
   \sI}}).
\end{eqnarray}
Evolution equations implied by this bracket are
\begin{subequations}\label{eq.PB.DA}
\begin{eqnarray}
	\partial_t \rho &=& -\partial_i(\rho E_{m_i})\\
	\partial_t m_i &=& -\rho\partial_i E_\rho - m_j \partial_i E_{m_j} -s \partial_i E_s - 
	\A{L}{l}\partial_i E_{\A{L}{l}} -D^l_{\ L} \partial_j E_{D^l_{\ L}}\nonumber\\
		&&+\partial_i(\A{L}{l} E_{\A{L}{l}}) - \partial_l(\A{L}{i} E_{\A{L}{l}}) +\partial_j(D^j_{\ 
		I} 
		E_{D^i_{\ \sI}})\\
	\partial_t \A{I}{i} &=& E_{D^i_{\ \sI}} -E_{m_j}\partial_j \A{I}{i} -\A{I}{j} \partial_i 
	E_{m_j}\nonumber\\
	 &=& E_{D^i_{\ \sI}} -\partial_i (\A{I}{j} E_{m_j}) + (\partial_i \A{I}{j} - \partial_j 
	 \A{I}{i}) 
	 E_{m_j}\\
	 \partial_t D^i_{\ \sI} &=& -E_{\A{I}{i}} -\partial_j(D^i_{\ \sI} E_{m_j})+D^j_{\ 
	 \sI}\partial_j 
	 E_{m_i}\\
	\partial_t s &=& -\partial_i(s E_{m_i}).
\end{eqnarray}
\end{subequations}
In order to have reversible dynamics, parity of $\DD$ with respect to time reversal must be odd 
(parity of $\AA$ is even). The vector field $\DD$ can be thus interpreted as the associated 
momentum of distortion (e.g. it may represent microinertia of the microstructure, e.g. see 
\cite{PRD-Torsion2019}). By letting the functional depend only on $(\rho,\mm,s,\AA)$ bracket 
\eqref{eq.PB.DA} becomes bracket \eqref{eq.PB.A}, and it can be thus seen as an extension of that 
bracket.

In summary, we have constructed the Hamiltonian dynamics of a cotangent bundle advected by fluid mechanics. Distortion $\AA$ is coupled to its associated momentum $\DD$ (similarly as in \cite{PRD-Torsion2019}), and the resulting equations are indeed Hamiltonian. In particular, by disregarding the associated momentum, the bracket for distortion matrix \eqref{eq.PB.A} can be seen as the Lie-Poisson bracket for semidirect product of fluid mechanics and a vector space, which is the geometric interpretation of the Poisson bracket. This means that the SHTC equations can be interpreted as evolution of total momentum advecting all the other fields ($\rho$, $s$, $\AA$). 

\subsection{Onsager-Casimir reciprocal relations}\label{sec.OCRR}
Let us now recall another feature of Hamiltonian mechanics, namely the Onsager-Casimir reciprocal relations.
We assume that the Hamiltonian evolution is time-reversible (at least short time, assuming the existence of strong solutions), which is usually the case. From the mathematical point of view time-reversibility follows directly from the construction of Poisson brackets on symplectic manifolds or from subsequent reductions towards less detailed levels of description, see \cite{PRE15}. Alternatively, as follows from our discussion in Section \ref{sec.gauge}, it is a consequence of energy conservation in Hamiltonian systems. From the physical point of view, Hamiltonian dynamics expresses mechanics, which constructed as reversible. Irreversibility appears once thermodynamic effects are taken into account. 

Assuming, moreover, that all state variables have definite parities with respect to the 
time-reversal transformation (TRT), 
\begin{equation}
\TRT(q^\alpha) = \Par(q^\alpha) q^\alpha, \qquad \Par(q^\alpha) = \pm 1.
\end{equation}
TRT inverts velocities of all particles and all velocity-like and momentum-like (generally quantities odd w.r.t. TRT) fields. For instance $\Par(x^i) = 1$, $\Par(m_i) = -1$, $\Par(s)=\Par(e)=1$, $\Par(\rho)=1$ and $\Par(\AA)=1$. For the Hamiltonian evolution to generate reversible evolution, the Poisson bivector must satisfy
\begin{equation}\label{eq.L.rev}
    \Par(L^{\alpha\beta}) = -\Par(q^\alpha)\Par(q^\beta),
\end{equation}
see \cite{PRE15}. Therefore, $\Par(L^{\alpha\beta})=1$ for $\Par(q^\alpha)=-\Par(q^\beta)$ while $\Par(L^{\alpha\beta})=-1$ for $\Par(q^\alpha)=-\Par(q^\beta)$.
Regarding the Hamiltonian evolution equations \eqref{eq.Ham.sys}, condition \eqref{eq.L.rev} 
together with antisymmetry of $\LL$, $L^{\alpha\beta}=-L^{\beta\alpha}$, imply the following theorem:
\begin{theorem}[Onsager-Casimir reciprocal relations]
Assuming Hamiltonian evolution \eqref{eq.Ham.sys} with reversible Poisson bivector (condition \eqref{eq.L.rev}), then 
\begin{itemize}
    \item variables with the same parity, $\Par(q^\alpha)=\Par(q^\beta)$, are coupled by an operator symmetric with respect to simultaneous transposition and time reversal
    \item variables with opposite parities, $\Par(q^\alpha)=-\Par(q^\beta)$, are coupled by an operator antisymmetric with respect to simultaneous transposition and time reversal.
\end{itemize}
See \cite{HCO,PKG}.
\end{theorem}
Onsager-Casimir reciprocal relations are thus automatically satisfied by reversible Hamiltonian evolution. Of course the irreversible part is also required to satisfy these relations, see \cite{HCO, PKG}. 

\section{Space-time formulation}\label{sec.Lagr.mech}
In this section, we present a variational formulation of the Hamiltonian continuum mechanics 
discussed 
in Sec.\,\ref{sec.Ham}. As it was mentioned in the Introduction, we are motivated here by the 
variational formulations of fluid mechanics \cite{Gavrilyuk1999,Gavrilyuk-variational} and space-time formulations of Hamiltonian continuum mechanics, which are
required for dealing with relativistic flows \cite{Ottinger1998,HCO,PTRSA2019}, and also by the possibility to
improve consistency of the structure-preserving numerical methods as the more general structure 
of the equations can be revealed. In particular, we shall see that the 
variational space-time formulation provides the evolutionary equations which are equivalent to 
those obtained from the bracket formulation, equations \eqref{eq.evo.Eu} and \eqref{eq.evo.A}. This 
may suggest a way to the generalization of the Hamiltonian formulation with distortion, discussed in 
Sec.\,\ref{sec.Ham}, to a space-time 
formulation.


The index convention used in this section is as follows. First letters of the Latin 
alphabet $ a,b,... 
$ and $ 
\sA,\sB,... $ run from $ 0 $ to $ 3 $  while middle letters $ i,j,... $ and $ I,J,... $ run from $ 
1 $ to $ 3 $. Also, small letters index quantities related to the Eulerian coordinates $ x^a $ 
while capital letters index quantities related to the Lagrangian coordinates $ X^\sA $.

\subsection{Lagrangian variational formulation}\label{sec.variation.Lagr}

Let us consider the mapping $x^a(X^\sA)$, with the inverse mapping $ X^\sA(x^a) $,  $a=0,1,2,3$, 
$\sA=0,1,2,3$ such that $x^0=t$ is the time of an Eulerian observer while $X^0 = \tau$ is the time 
of a Lagrangian observer (co-moving with the medium). However, we shall assume 
non-relativistic settings 
and hence, $ t = \tau $. 

Let us first consider a general Lagrangian density $ \tilde{\Lambda}(X^\sA,x^a(X^\sA),\pd_\sA x^a) 
$ in the 
Lagrangian 
frame $ X^\sA $, where $ \pd_\sA = \frac{\pd}{\pd X^\sA} $. We assume that $ 
\tilde{\Lambda} $ depends only on the 4-potential 
$ x^a(X^\sA) $ and their first derivatives. Moreover, due to that the Lagrangian should transform 
as a proper scalar with respect to the Galilean group of transformations, $ \tilde{\Lambda} $ 
depends on $ X^\sA $ and $ x^a(X^\sA) $ 
only implicitly, i.e.
$ \tilde{\Lambda}(X^\sA,x^a(X^\sA),\pd_\sA x^a) = \Lambda(\pd_\sA x^a) $.
 The action then reads
\begin{equation}
S = \int \tilde{\Lambda}(X^\sA,x^a(X^\sA),\pd_\sA x^a)\dX = \int \Lambda(\pd_\sA x^a)\dX,
\end{equation}
variation of which with respect to $ \delta x^a $ gives the Euler-Lagrange equation
\begin{equation}\label{eqn.EL.Lagr}
\pd_\sA \Lambda_{\pd_{\sA} x^a} = 0.
\end{equation}
This equation governs motion of the continuum,
and is equivalent to \eqref{eq.Ene.L}, \eqref{eq.L.evo.fin}. Equation of motion 
\eqref{eqn.EL.Lagr} 
is
a system of second-order PDEs on $ x^a $. It, however, can be seen as a first order system on 
4-deformation gradient $ 
\F{a}{A} = \pd_\sA x^a$ supplemented by the integrability condition 
$ \pd_\sA \F{a}{B} - \pd_\sB \F{a}{A} = 0$,
which is a trivial consequence of the definition $ \F{a}{A} = \pd_\sA x^a $ and 
the continuity of $ x^a(X^\sA) $. Thus, the first-order 
system equivalent to \eqref{eqn.EL.Lagr} and \eqref{eq.Ene.L}, \eqref{eq.L.evo.fin} is
\begin{equation}\label{eqn.EL.Lagr.extend}
\pd_\sA \Lambda_{\F{a}{A}} = 0, \qquad \pd_\sA \F{a}{B} - \pd_\sB \F{a}{A} = 0.
\end{equation}

Recall that the 4-velocity is conventionally defined as the tangent vector to the continuum 
particle trajectories
\begin{equation}
    u^a = \frac{\pd x^a}{\pd \tau} = \frac{\pd x^a}{\pd X^0}  = \F{a}{0},
\end{equation}
which gives $ u^a = (1,v^1,v^2,v^3) $ with $ v^i = \frac{\pd x^i}{\pd \tau} $ being the ordinary 
3-velocity. Therefore, the 4-velocity $ u^a $ is just the $ 0 $-th column of the 4-deformation 
gradient $ \F{a}{\sA} $ \cite{PRD-Torsion2019}. Hence, equations 
\eqref{eqn.EL.Lagr.extend}$ _1 $ are essentially equations for $ u^a $.

\subsection{Eulerian variational formulation}\label{sec.variation.Euler}
Alternatively, the action integral can be transformed into the Eulerian frame. For 
this purpose, it is sufficient to perform the change of variables $ X^\sA \rightarrow 
x^a(X^\sA) $ 
in the action \eqref{eqn.EL.Lagr}:
\begin{eqnarray}\label{eq.Action.transform}
S &=& \int \tilde{\Lambda}(X^\sA,x^a(X^\sA),\pd_\sA x^a)\dX \nonumber\\
  &=& \int w \, \tilde{L}(X^\sA(x^a),x^a,\pd_a X^\sA)\dx \nonumber\\
  &=& \int \tilde{\mathcal{L}}(X^\sA(x^a),x^a,\pd_a X^\sA)\dx
\end{eqnarray}
where $ w = \det(\pd_a X^\sA) $, $ \tilde{\Lambda}(\pd_\sA x^a) = \tilde{L}(\pd_a X^\sA) $, and $ 
\tilde{\mathcal{L}} = w \tilde{L} $. The Euler-Lagrange 
equation is
\begin{equation}\label{EL.Euler}
\pd_a \mathcal{L}_{\pd_a X^\sA} = 0,
\end{equation}
where $ \mathcal{L}(\pd_a X^\sA) = \tilde{\mathcal{L}}(X^\sA(x^a),x^a,\pd_a X^\sA) $.
Similarly as in \eqref{eqn.EL.Lagr.extend}, it can be viewed as an extended first-order system for $ 
\A{A}{a} = \pd_a X^\sA $,
\begin{equation}\label{Euler.A}
\pd_a (\A{A}{b}\mathcal{L}_{\A{A}{a}} - \mathcal{L}\delta^a_{\ b}) = 0, \qquad u^a(\pd_a \A{A}{b} - 
\pd_b \A{A}{a}) = 0,
\end{equation}
where the former equation is obtained from \eqref{EL.Euler} using the fact that $ \pd_a\A{A}{b} = 
\pd_b \A{A}{a} $ while the 
latter is the result of transformation of \eqref{eqn.EL.Lagr.extend}$ _2 $ into the 
Eulerian frame. The tensor $ T^a_{\ \,b} = \A{A}{b}\mathcal{L}_{\A{A}{a}} - \mathcal{L}\delta^a_{\ 
b} 
$ can be naturally called the matter \textit{energy-momentum tensor}.

We note that in the computational elasticity the numerical solution should fulfill the 
compatibility condition $ \partial_i \A{I}{j} - \partial_j \A{I}{i} = 0 $ which cannot be 
guarantied if a general purpose numerical method is used (especially in the long time simulations).
Instead, a curl-preserving numerical integrator have to be used, e.g. 
\cite{JeltschTorrilhon2006,Torrilhon2004,DumbserCurl2020}. It is therefore important to understand 
while 
constructing such a method that the whole evolution equation for the distortion field has the 
structure of the four-curl. This fact can be used in order to chose a proper time integration which 
is one of the critical steps in designing structure preserving numerical integrators.

\subsection{Equivalence of the variational and Hamiltonian formulations}

Let us now show that the Eulerian equations \eqref{eq.evo.Eu.F} and \eqref{eq.evo.Eu.A} for $ 
\F{i}{I} $ and $ \A{I}{i} $ can be derived from the space-time formulation 
\eqref{eqn.EL.Lagr.extend}. We first note that the Lagrangian equation for $ \F{a}{A} $ is just 
an identity, so-called \textit{integrability condition},
\begin{equation}\label{eqn.F.4d}
\pd_\sA (\pd_\sB x^a) - \pd_\sB (\pd_\sA x^a) = \pd_\sA \F{a}{B} - \pd_\sB \F{a}{A} = 0.
\end{equation}
Thus, if we consider the $ 0 $-th ($ \sA=0 $) equation 
and 
use that $ v^i = \F{i}{0} $, we obtain the conventional time evolution equation for the spatial 
entries of $ \F{a}{\sA} $
\begin{equation}\label{eqn.F.3d}
\pd_\tau \F{i}{I} - \pd_I v^i = 0.
\end{equation}

One can easily get an Eulerian form of this PDE
\begin{equation}\label{eqn.F.euler}
\pd_t \F{i}{I} + v^k\pd_k \F{i}{I} - \F{k}{I}\pd_k v^i = 0
\end{equation}
by simply transforming the Lagrangian derivatives to Eulerian ones $ \pd_\sA = \F{a}{\sA}\pd_a $ 
(and in particular $ \pd_\tau = \F{a}{0}\pd_a = u^a \pd_a = \pd_t + v^k\pd_k $).
Similarly, \eqref{Euler.A} reduces to\footnote{One should use the identity $ u^a\pd_b\A{A}{a} = 
-\A{A}{a}\pd_b u^a$, and the so-called orthogonality condition $ u^a\A{I}{a} = \F{a}{0}\A{I}{a} = 0 
$ and hence $ \A{I}{0} = -v^k \A{I}{k} $.}
\begin{equation}\label{eqn.A.3d}
\pd_t \A{I}{i} + v^k \pd_k \A{I}{i} + \A{I}{k}\pd_i v^k = 0.
\end{equation}
Equations \eqref{eqn.F.euler} and \eqref{eqn.A.3d} are exactly equations \eqref{eq.evo.Eu.F} and 
\eqref{eq.evo.Eu.A}, correspondingly, generated by the corresponding Poisson brackets.

Having two Eulerian equations at hand, \eqref{eqn.F.euler} and \eqref{eqn.A.3d}, which one can be considered as a preferable equation to be used in the Eulerian frame.

Although the equations are 
equivalent (at least regarding strong solutions), the variational formulation suggests that $ \F{a}{A} $ is a natural 
state variable in the Lagrangian frame while $ \A{A}{a} $ is naturally to be used in the Eulerian 
frame, see action integrals \eqref{eq.Action.transform}. 
 
Furthermore, in order to see more differences between these two equations, let us write 
them in equivalent \textit{semi-conservative} forms. 
Thus, let us introduce $ \hatF{i}{I} = \rho \F{i}{I} $ like in 
works~\cite{GodRom2003,Markus2009} with $ \rho = \rho_0/\det(\F{i}{I})$ being the mass density, and 
$ \rho_0 $ the reference mass density. Then, by multiplying \eqref{eqn.F.euler} by $ \rho $ and 
adding 
it with 
the continuity equation multiplied by $ \F{i}{I} $, one gets
\begin{equation}\label{eqn.F.semicons}
\pd_t\hatF{i}{I} + \pd_k (v^k \hatF{i}{I} - v^i \hatF{k}{I}) + v^i \pd_k\hatF{k}{I} = 0.
\end{equation}
On the other hand, by adding $ 0 \equiv v^k\pd_i\A{I}{k} - v^k\pd_i\A{I}{k} $ to the equation for $ 
\A{I}{i} $, it can be written as
\begin{equation}\label{eqn.A.semicons}
\pd_t \A{I}{i} + \pd_i (v^k \A{I}{k}) + v^k (\pd_k \A{I}{i} - \pd_i \A{I}{k})  = 0.
\end{equation}
A few remarks are in order, which discuss similarities and differences between formulations 
in terms of $ \F{a}{A} $ and $ \A{A}{a} $.

\begin{remark}\label{rem.div.curl}
The last terms in \eqref{eqn.F.semicons} and \eqref{eqn.A.semicons} are 
connected as
\begin{equation}
\pd_k \hatF{k}{I} = \rho \F{i}{I}\F{k}{J}(\pd_k \A{J}{i} - \pd_i \A{J}{k}).
\end{equation}
In particular, $ \pd_k \A{J}{i} - \pd_i \A{J}{k} = 0 $ in elasticity and hence, $ 
\pd_k\hatF{k}{I} = 0 $ as well. However, one should not remove these terms from the PDEs because 
the 
resulting equations would not be equivalent to the original ones. In 
particular, the removing of these terms changes the characteristic structure of 
the equations \cite{SHTC-GENERIC}. 
\end{remark}


\begin{remark} In the space-time formulation, the equations for $ \F{a}{A} $ 
and for $ 
\A{A}{a} $ are nothing else but the Lie derivatives along the 4-velocity $ u^a = \F{a}{0}$:
\begin{equation}\label{eq.Lie.F.A}
\Lie_{\bm{u}} \F{a}{A} = u^b\pd_b \F{a}{A} - \F{b}{A}\pd_b u^a = 0, \qquad 
\Lie_{\bm{u}} \A{A}{a} = u^b \pd_b \A{A}{a} + \A{A}{b}\pd_a u^b = 0.
\end{equation}
Here, it is necessary to recall that $ \F{a}{A} $ and $ \A{A}{a} $ transform not as rank-2 
space-time tensors 
but as tetrads of covariant and contravariant vectors correspondingly, see also 
Sec.\,\ref{sec.SP}. Using 
this fact, the ordinary partial derivatives $ \pd_a $ in equations \eqref{eq.Lie.F.A} can be 
replaced with the covariant derivatives $ \nabla_a $ associated with the torsion-free Levi-Civita 
connection:
\begin{equation}\label{eq.Lie.F.A.cov}
u^b\nabla_b \F{a}{A} - \F{b}{A}\nabla_b u^a = 0, \qquad 
u^b \nabla_b \A{A}{a} + \A{A}{b}\nabla_a u^b = 0.
\end{equation}
This means that both formulations are unconditionally covariant, that is they 
transform form-invariantly under general coordinate transformation (including 
transformations between non-inertial frames).
Furthermore, as discussed in \cite{Frewer2009}, because the time evolutions 
\eqref{eq.Lie.F.A.cov} are given by the Lie derivatives, they are intrinsically 
transformed as objective tensors, that is frame-independently, and moreover, 
they 
are defined unambiguously. This, in particular, makes such formulations 
attractive 
for using in rheology where the classical stress-based formulations (e.g. 
Maxwell's viscoelastic model) are known to suffer 
from the non-uniqueness of the choice of objective time rates.
\end{remark}

\begin{remark}
The four-dimensional equations for $ \F{a}{A} $ and $ \A{A}{a} $ are related as
\begin{equation}
\pd_\sA \F{c}{B} - \pd_\sB \F{c}{A} = -\F{a}{A}\F{b}{B}\F{c}{C}(\pd_a\A{C}{b} - \pd_b\A{C}{a}).
\end{equation}
Hence, in the elasticity settings, formulations in terms of $ \F{a}{A} $ and $ \A{A}{a} $ can be 
used interchangeably, and, we believe, there is no universal way to give a 
preference to one over another. 
However, the situation becomes different in the context of irreversible deformations of either 
solids or fluids, and $ \A{A}{a} 
$-formulation may give more insight into the non-elastic phenomena via the concepts of 
torsion $ 
\mathfrak{T}^{\sA}_{\ ab} = \pd_a\A{A}{b} - \pd_b\A{A}{a} $ \eqref{torsion}, non-holonomic tetrads, 
and non-Riemannian geometry, e.g see \cite{PRD-Torsion2019,Yavari2012}.
\end{remark}

\section{Conclusion}
Continuum mechanics can be constructed from the principle of 
least action or as Hamiltonian mechanics. In the Lagrangian frame (attached to matter) it has the symplectic structure of Hamilton canonical equations.
In the Eulerian frame (attached to an inertial system), 
however, it only has Poisson structure and the underlying Poisson brackets are more complicated.
Here, we derive the Poisson bracket for density, momentum density, entropy density and deformation 
gradient (or distortion) by reduction of the Lagrangian frame to the Eulerian frame, which 
generalizes earlier results from the literature. We also show how to derive an even more general 
bracket equipped with extra mass flux.

It requires some time to learn how to handle Poisson brackets. So why should one pay attention to them? Features of Hamiltonian mechanics include:
\begin{itemize}
    \item Clear geometric origin, Sec. \ref{sec.Euler}.
    \item Automatic energy conservation, Sec. \ref{sec.Ham}.
    \item Automatic relation to hyperbolicity, see Sec. \ref{sec.hype}.
    \item Geometric self-consistency, see Sec. \ref{sec.Jacobi}.
    \item Relation between gauge invariance and conservation laws, see Sec. \ref{sec.gauge}.
    \item Variational principles, see Sec. \ref{sec.Clebsch} and \ref{sec.variation.Lagr}.
    \item Underlying structure of Lie groups (semidirect product), see Sec. \ref{sec.SP}.
    \item Automatic validity of Onsager-Casimir reciprocal relations, see Sec. \ref{sec.OCRR}.
    \item Robustness with respect to violation of deformation compatibility conditions 
(propagation of torsion tensor), sec. \ref{sec.SHTC}.
\end{itemize}

\section*{Acknowledgment}
This research was originally initiated by Markus H{\" u}tter during the IWNET 
2018 conference. M.P. is grateful to Miroslav Grmela, Petr Vágner, O{\u g}ul 
Esen and V{\' i}t Pr{\r u}{\v s}a for many discussions on Hamiltonian 
formulation of solids and fluids and to Jan Zeman, Petr Pelech, Martin Sýkora, 
Miroslav Hanzelka and Tomáš Los for patience and their opinions during the 
GENERIC course 2018/19. Especially Petr Pelech and Martin S{\' y}kora helped us 
checking a preliminary version of the manuscript.

I.P. acknowledges a financial support by Agence Nationale de la Recherche (FR) (Grant No. 
ANR-11-LABX-0040-CIMI) within the program ANR-11-IDEX-0002-02. Also, 
I.P. acknowledges the financial support
received from the Italian Ministry of Education, University
and Research (MIUR) in the frame of the Departments of
Excellence Initiative 2018–2022 attributed to DICAM of the
University of Trento (grant L. 232/2016) and in the frame
of the PRIN 2017 project. I.P. has further received funding
from the University of Trento via the UniTN Starting
Grant initiative.
 M.P. and V.K. were supported by Czech Science Foundation, Project No. 17-15498Y, and by Charles 
 University Research Program No. UNCE/SCI/023. 




\appendix
\section{Functional derivatives}\label{sec.FD}
The purpose of this section is to recall the concept of functional derivative. 
Consider a functional $\Ffunc$ of field $f(\XX)$ that is Fréchet 
differentiable, i.e. 
\begin{equation}
	\Ffunc(f+\delta f) = \Ffunc(f) +  D \Ffunc|_f (\delta f) + \OBig(\delta 
	f)^2,
\end{equation}
where $\delta f \in C^\infty_0$ is a smooth variation with compact support (zero at the boundaries). Topology can be specified for instance as that of $\mathcal{D}$ space \cite{Schwartz}.
The Fréchet differential is then equal to the Gateaux derivative
\begin{equation}
	\frac{\diff}{\diff \lambda}\Big|_{\lambda = 0} \Ffunc(f+\lambda \delta f) = 
	D \Ffunc|_f (\delta f).
\end{equation}
Since the Fréchet differential is linear in its argument $(\bullet)$, it can be seen as an element of the dual space to $\delta f$, which is the space of distributions $\mathcal{D}'$. Therefore, it can be represented by 
\begin{equation}
	D \Ffunc_f(\delta f) = \int\dX \frac{\delta \Ffunc}{\delta f} \delta f,
\end{equation}
where the integral is understood as a notational shorthand for duality in the distributional sense $\langle\bullet,\bullet\rangle$. By combining the last two equalities, we obtain
\begin{equation}
	\frac{\diff}{\diff \lambda}\Big|_{\lambda = 0} \Ffunc(f+\lambda \delta f) 
	=  \int\dX \frac{\delta \Ffunc}{\delta f} \delta f,
\end{equation}
which is the usual way for calculation of functional derivatives $\frac{\delta 
\Ffunc}{\delta f}$.

For instance if $\Ffunc$ is integral of a smooth real-valued function $g(f)$ of 
field $f(\XX)$, the functional derivative becomes
\begin{equation}
	\frac{\delta}{\delta f}\int \dX g(f(\XX)) = g'(f(\XX)),
\end{equation}
which is just the ordinary derivative of $g$. 

If the functional depends on gradient of $f$, we have to carry out integration by parts (recalling that $\delta f$ vanishes at the boundaries), e.g.
\begin{equation}
	\frac{\delta}{\delta f}\int \dX \frac{1}{2}\nabla_\XX f(\XX) \cdot \nabla_\XX f(\XX) = -\nabla_\XX \cdot (\nabla_\XX f).
\end{equation}

Finally, if the functional simply picks a value of the field at given position, its derivative is the Dirac $\delta-$distribution, 
\begin{equation}
\Ffunc(f) = f(\XX_0) = \int\diff \XX \delta(\XX-\XX_0)f(\XX) \Rightarrow 
\frac{\delta \Ffunc}{\delta f(\XX)} = \delta(\XX-\XX_0).
\end{equation}
Equipped with these instruments, we can approach the transformation of Poisson brackets from the Lagrangian frame to the Eulerian frame.

\section{From Lagrange to Euler}\label{sec.L-E}
The purpose of this rather technical Appendix is to show in detail how the Eulerian bracket \eqref{eq.PB.Eu} is obtained from the Lagrangian canonical bracket \eqref{eq.PB.L}. The latter bracket expresses kinematics of fields $\xx(\XX)$ and $\MM(\XX)$, while the former bracket has only Eulerian state variables $\rho(\xx)$, $\mm(\xx)$, $s(\xx)$ and $\FF(\xx)$. Note that the calculations are purely formal as we do not discuss the analytical details of the studied objects, which are thus assumed to be regular enough.

\begin{figure}
    \centering
    \includegraphics[scale=0.3]{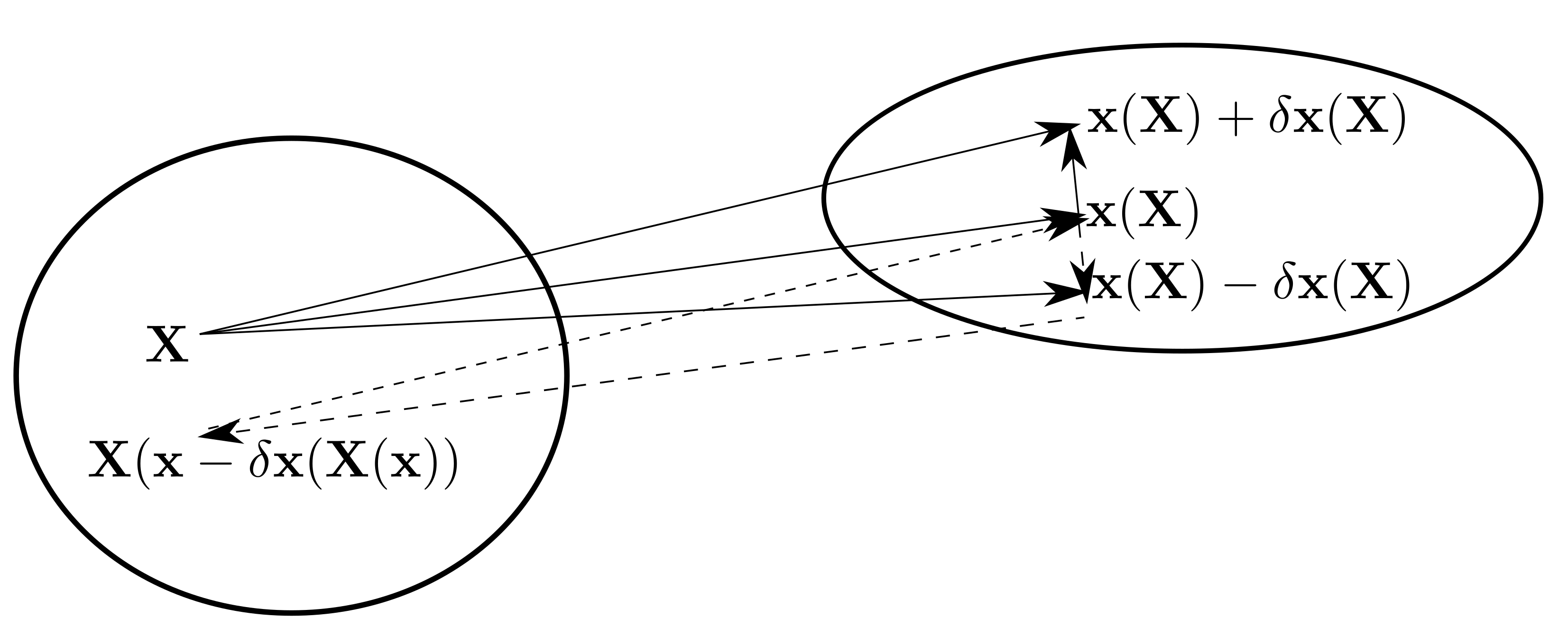}
    \caption{Demonstration of formula \eqref{eq.xX} (the dashed triangle). The left manifold represents the Lagrangian configuration while right represents the Eulerian configuration. }
    \label{fig.trafo}
\end{figure}

Before carrying out the actual transformation, we make a few observations about the mapping $\xx(\XX)$ and its inverse $\XX(\xx)$ and their behavior with respect to perturbations $\delta \xx(\XX)$. Firstly, we see that
\begin{multline}\label{eq.xX}
	(\xx(\XX) + \delta\xx(\XX))\circ \XX(\xx-\delta \xx(\XX(\xx))) \\
	= \xx -\delta\xx(\XX(\xx)) + \delta\xx(\XX(\xx)) + \OBig(\delta\xx)^2
= \xx + \OBig(\delta\xx)^2,
\end{multline}
which helps when calculating functional derivatives with respect to $\xx(\XX)$. This identity is demonstrated on Fig. \ref{fig.trafo}.

The functional derivatives of the Eulerian fields with respect to the Lagrangian fields are necessary to perform the reduction. Let us start with the Lagrangian density in the Eulerian frame, $\rho_0(\xx)\stackrel{def}{=}\rho_0(\XX(\xx))$. This slightly overloaded notation is used throughout this appendix. To avoid confusion, the spatial variables on which the fields depend will be always indicated. To find functional derivative of $\rho_0(\xx)$ with respect to $\xx(\XX)$ we study 
\begin{equation}
	\rho_0(\xx;\xx+\delta\xx)\stackrel{def}{=} \rho_0(\XX(\xx)+\delta\XX(\xx)),
\end{equation}
where $\delta\XX(\xx)$ is the perturbation of mapping $\XX(\xx)$ induced by perturbation $\delta\xx(\XX)$. Using relation \eqref{eq.xX}, we obtain 
\begin{eqnarray}
	\rho_0(\xx;\xx+\delta\xx) &=& \rho_0(\XX(\xx-\delta\xx(\XX(\xx)))) = \rho_0(\xx-\delta\xx(\XX(\xx)))\\
	&=&\rho_0(\xx)-\partial_k \rho_0(\xx) \delta x^k(\XX(\xx))+\OBig(\delta\xx)^2\nonumber\\
	&=&\rho_0(\xx)+\int\dX \left(-\partial_k \rho_0(\xx) \delta(\XX-\XX(\xx))\right)\delta x^k(\XX)+\OBig(\delta\xx)^2,\nonumber
\end{eqnarray}
which means that
\begin{equation}\label{eq.rho0x}
	\frac{\delta \rho_0(\xx)}{\delta x^k(\XX)} = -\partial_k \rho_0(\xx) \delta(\XX-\XX(\xx)).
\end{equation}

\subsection{Derivative of the Eulerian deformation gradient $\FF(\xx)$}
The Eulerian mass density $\rho(\xx)$ is equal to $\rho_0(\xx)$ multiplied by Jacobian of the mapping $\XX(\xx)$, and so to acquire the functional derivative of $\rho(\xx)$ we have to first  deal with functional derivative of the Eulerian deformation gradient
\begin{equation}
	\FF(\xx)\stackrel{def}{=}\frac{\partial \xx}{\partial \XX}\Big|_{\XX(\xx)}.
\end{equation}
Using again relation \eqref{eq.xX} we have
\begin{eqnarray}
	\F{i}{I}(\xx;\xx+\delta\xx) &\stackrel{def}{=}&\frac{\partial x^i(\XX) +\delta 
	x^i(\XX)}{\partial X^I}\Big|_{\XX(\xx-\delta\xx(\XX(\xx)))}\\
	&=&\frac{\partial x^i}{\partial X^I}\Big|_{\XX(\xx-\delta\xx(\XX(\xx)))} + \frac{\partial \delta x^i}{\partial X^I}\Big|_{\XX(\xx)}+\OBig(\delta\xx)^2\nonumber\\
	&=&\int\dX\delta(\XX-\XX(\xx)) \frac{\partial \delta x^i}{\delta X^I} 
	+\underbrace{\frac{\partial x^i}{\partial X^I}\Big|_{\XX(\xx)}}_{=\F{i}{I}(\xx)} \nonumber\\
	&&-\partial_k 
	\F{i}{I}(\xx)\delta x^k(\XX(\xx)) + \OBig(\delta\xx)^2\nonumber\\
	&=&\F{i}{I}(\xx) - \int\dX\frac{\partial \delta(\XX-\XX(\xx))}{\partial X^I}\delta x^i(\XX) \nonumber\\
	&&-\int\dX\partial_k \F{i}{I}(\xx)\delta(\XX-\XX(\xx))\delta x^k(\XX),\nonumber
\end{eqnarray}
which means that the sought functional derivative reads
\begin{equation}
	\frac{\delta \F{i}{I}(\xx)}{\delta x^k(\XX)} = 
	- \frac{\partial \delta(\XX-\XX(\xx))}{\partial X^I}\delta^i_k -\partial_k 
	\F{i}{I}(\xx)\delta(\XX-\XX(\xx)).
\end{equation}

\subsection{Derivative of the Eulerian mass density $\rho(\xx)$}
Now we can finish the calculation of the functional derivative of the Eulerian density $\rho(\xx)$ with respect to $\xx(\XX)$. The first part, derivative of $\rho_0(\xx)$, has already been obtained before in Eq. \eqref{eq.rho0x}. What remains is to calculate derivative of determinant $\det \FF(\xx)$ with respect to $\xx(\XX)$. 

Considering determinant of a general matrix $\CC$, its variation when the matrix is perturbed by $\delta\CC$ reads
\begin{multline}
	\det(\CC+\delta\CC) = \det(\CC)\cdot\det(\Id+\CC^{-1}\cdot\delta\CC) \\
	= \det \CC + \det\CC\cdot\Tr(\CC^{-1}\cdot\delta\CC) + \OBig(\delta \CC)^2.
\end{multline}
Therefore, derivative of $\det(\FF)$ is
\begin{equation}
	\frac{\delta \det \FF^{-1}(\xx)}{\delta x^k(\XX)}
	= -\frac{1}{(\det \FF(\xx))^2} \det \FF(\xx) \frac{\partial X^I}{\partial x^i} \frac{\delta 
	\F{i}{I}(\xx)}{\delta x^k(\XX)}.
\end{equation}

Derivative of the Eulerian density finally reads
\begin{eqnarray}
	\frac{\delta \rho(\xx)}{\delta x^k(\XX)}&=&\frac{\delta \rho_0(\xx)}{\delta x^k(\XX)} \det \frac{\partial \XX}{\partial \xx}
	-\frac{\rho_0(\xx)}{(\det \FF(\xx))^2} \det \FF(\xx) \frac{\partial X^I}{\partial x^i} 
	\frac{\delta \F{i}{I}(\xx)}{\delta x^k(\XX)}\nonumber\\
	&=&-\partial_k \rho_0(\xx) \delta(\XX-\XX(\xx))\frac{1}{\det\FF(\xx)}\nonumber\\
	&&
	+\frac{\rho_0(\xx)}{\det \FF(\xx)} \frac{\partial X^I}{\partial x^i} \frac{\partial \delta(\XX-\XX(\xx))}{\partial X^I}\delta^i_k \nonumber\\
	&&+\frac{\rho_0(\xx)}{\det \FF(\xx)} \frac{\partial X^I}{\partial x^i} \partial_k 
	\F{i}{I}(\xx)\delta(\XX-\XX(\xx)).
\end{eqnarray}

\subsection{Derivative of the Eulerian entropy density $s(\xx)$}
Having calculated derivative of mass density, the result for entropy density $s(\xx)=s_0(\xx)/\det\FF(\xx)$  is analogous,
\begin{eqnarray}
	\frac{\delta s(\xx)}{\delta x^k(\XX)}&=&
	-\partial_k s_0(\xx) \delta(\XX-\XX(\xx))\frac{1}{\det\FF(\xx)}\nonumber\\
	&&
	+\frac{s_0(\xx)}{\det \FF(\xx)} \frac{\partial X^I}{\partial x^i} \frac{\partial \delta(\XX-\XX(\xx))}{\partial X^I}\delta^i_k \nonumber\\
	&&+\frac{s_0(\xx)}{\det \FF(\xx)} \frac{\partial X^I}{\partial x^i} \partial_k 
	\F{i}{I}(\xx)\delta(\XX-\XX(\xx)).
\end{eqnarray}

\subsection{Derivative of the Eulerian momentum density $\mm(\xx)$}
The functional derivative of $\mm(\xx)=\MM(\XX(\xx))/\det \FF(\xx)$ with respect to $\xx(\XX)$ has the same form as derivatives of $\rho(\xx)$ and $s(\xx)$,
\begin{eqnarray}
	\frac{\delta m_l(\xx)}{\delta x^k(\XX)}&=&
	-\partial_k M_l(\xx) \delta(\XX-\XX(\xx))\frac{1}{\det\FF(\xx)}\nonumber\\
	&&
	+\frac{M_l(\xx)}{\det \FF(\xx)} \frac{\partial X^I}{\partial x^i} \frac{\partial \delta(\XX-\XX(\xx))}{\partial X^I}\delta^i_k \nonumber\\
	&&+\frac{M_l(\xx)}{\det \FF(\xx)} \frac{\partial X^I}{\partial x^i} \partial_k 
	\F{i}{I}(\xx)\delta(\XX-\XX(\xx)).
\end{eqnarray}

But the field $\mm(\xx)$ also depends on the yet unused Lagrangian field $\MM(\XX)$. Derivative with respect to this fields is
\begin{eqnarray}\label{eq.mM}
	\frac{\delta m_l(\xx)}{\delta M_k(\XX)} = \delta^k_l \frac{\delta(\XX-\XX(\xx))}{\det\FF(\xx)},
\end{eqnarray}
as follows from the formulas 
\begin{equation}
    \MM(\XX(\xx)) = \int\dX \delta(\XX-\XX(\xx)) \MM(\XX)
\end{equation}
and
\begin{equation}
    \frac{\delta M_k(\XX(\xx)}{\delta M_l(\XX)} = \delta^l_k \delta(\XX-\XX(\xx)).
\end{equation}

\subsection{Derivative of an arbitrary Eulerian functional}
Derivative of an arbitrary smooth enough functional\footnote{Functionals $F$ and $G$ are not used here to avoid confusing with tensor $\FF$.} of the Eulerian fields $C(\rho(\xx), \mm(\xx), s(\xx), \FF(\xx))$ with respect to the Lagrangian field $\xx(\XX)$ can be calculated by chain rule as
\begin{equation}\label{eq.Cx}
	\frac{\delta C}{\delta x^k(\XX)} = \int\dx \left(\frac{\delta C}{\delta \rho(\xx)}\frac{\delta \rho(\xx)}{\delta x^k(\XX)}
	+\frac{\delta C}{\delta m_l(\xx)}\frac{\delta m_l(\xx)}{\delta x^k(\XX)}
	+\frac{\delta C}{\delta s(\xx)}\frac{\delta s(\xx)}{\delta x^k(\XX)}
	+\frac{\delta C}{\delta \F{i}{I}(\xx)}\frac{\delta \F{i}{I}(\xx)}{\delta x^k(\XX)}\right).
\end{equation}
Similarly derivative of an arbitrary functional $D(\rho(\xx), \mm(\xx), s(\xx), \FF(\xx))$ with respect to the Lagrangian $\MM(\XX)$ field is
\begin{equation}
	\frac{\delta D}{\delta M_k(\XX)} = \int\dx \frac{\delta D}{\delta m_l(\xx)}\frac{\delta m_l(\xx)}{\delta M_k(\XX)},
\end{equation}
which, using Eq. \eqref{eq.mM}, can be rewritten more explicitly as
\begin{eqnarray}
	\frac{\delta D}{\delta M_k(\XX)} &=& \int\dx \frac{\delta D}{\delta m_l(\xx)}\delta^k_l \frac{\delta(\XX-\XX(\xx))}{\det\FF(\xx)}\nonumber\\
	&=& \int\dx \frac{\delta D}{\delta m_k(\xx)}\delta(\XX-\XX(\xx))\det\frac{\partial \XX}{\partial \xx}.
\end{eqnarray}
The $\delta-$distribution can be seen as the limit of a sequence of smooth functions (e.g. Gaussians), $f_n(\xx)\stackrel{\mathcal{D}'}{\to}\delta(\xx)$. Therefore, the last integral can be rewritten as
\begin{eqnarray}
	\frac{\delta D}{\delta M_k(\XX)} &=& \lim_{n\to\infty}\int\dx \frac{\delta D}{\delta m_k(\xx)}f_n(\XX-\XX(\xx)) \det\frac{\partial \XX}{\partial \xx}\nonumber\\
	&=& \lim_{n\to\infty}\int\dX' \frac{\delta D}{\delta m_k(\xx)}\Big|_{\xx(\XX')} f_n(\XX-\XX(\xx(\XX'))) \nonumber\\
	&=& \int\dX' \frac{\delta D}{\delta m_k(\xx)}\Big|_{\xx(\XX')} \lim_{n\to\infty}f_n(\XX-\XX') \nonumber\\
	&=& \int\dX' \frac{\delta D}{\delta m_k(\xx)}\Big|_{\xx(\XX')} \delta(\XX-\XX') \nonumber\\
	&=&\frac{\delta D}{\delta m_k(\xx)}\Big|_{\xx(\XX)}.
\end{eqnarray}
Now we are finally in position to calculate the Lagrangian Poisson bracket Eq. \eqref{eq.PB.L} for the Eulerian functionals $C$ and $D$. 

\subsection{The Eulerian Poisson bracket}
Bracket \eqref{eq.PB.L} is the sum of terms like
\begin{equation}
	\int\dX \frac{\delta C}{\delta x^k(\XX)} \frac{\delta D}{\delta M_k(\XX)},
\end{equation}
where the former functional derivative consists of all terms in Eq. \eqref{eq.Cx}. Let us first take only the term with derivative $C_{\rho(\xx)}$,
\begin{eqnarray}
	\int\dX \int\dx \frac{\delta C}{\delta \rho(\xx)}&&
	\left[
		-\partial_k \rho_0(\xx) \delta(\XX-\XX(\xx))\frac{1}{\det\FF(\xx)}\right.\nonumber\\
	&&
	+\frac{\rho_0(\xx)}{\det \FF(\xx)} \frac{\partial X^I}{\partial x^i} \frac{\partial \delta(\XX-\XX(\xx))}{\partial X^I}\delta^i_k \nonumber\\
	&&\left.+\frac{\rho_0(\xx)}{\det \FF(\xx)} \frac{\partial X^I}{\partial x^i} \partial_k 
	\F{i}{I}(\xx)\delta(\XX-\XX(\xx))\right]
\frac{\delta D}{\delta m_k(\xx)}\Big|_{\xx(\XX)}\nonumber\\
	&=& \int\dx \frac{\delta C}{\delta \rho(\xx)}\left[
		-\partial_k \rho_0(\xx)\det\FF^{-1}(\xx)\frac{\delta D}{\delta m_k(\xx)} \right.\nonumber\\
		&&-\rho_0(\xx) \partial_k \det \FF^{-1}(\xx)\frac{\delta D}{\delta m_k(\xx)} \nonumber\\
		&&\left.-\frac{\rho_0(\xx)}{\det \FF(\xx)} \frac{\partial X^I}{\partial x^i} \int\dX \delta(\XX-\XX(\xx))\frac{\partial}{\partial X^I}\frac{\delta D}{\delta m_i(\xx)}\Big|_{\xx(\XX)}\right]\nonumber\\
	&=&\int\dx \frac{\delta C}{\delta \rho(\xx)}\left[-\partial_k \rho(\xx)\frac{\delta D}{\delta m_k(\xx)}\right.\nonumber\\
	&&\left.\qquad -\rho(\xx) \partial_i \frac{\delta D}{\delta m_i(\xx)}\right]\nonumber\\
	&=& \int\dx \rho(\xx)\partial_i \frac{\delta C}{\delta \rho(\xx)} \frac{\delta D}{\delta m_i(\xx)},
\end{eqnarray}
which is obviously a part of the final Eulerian Poisson bracket \eqref{eq.PB.Eu}. In the same fashion we obtain
\begin{equation}
	\int\dX \int\dx \frac{\delta C}{\delta m_l(\xx)} \frac{\delta m_l(\xx)}{\delta x^k(\XX)} \frac{\delta D}{\delta M_k(\XX)}
	= \int\dx m_i(\xx)\partial_j \frac{\delta C}{\delta m_i(\xx)} \frac{\delta D}{\delta m_j(\xx)}
\end{equation}
and
\begin{equation}
	\int\dX \int\dx \frac{\delta C}{\delta s(\xx)} \frac{\delta s(\xx)}{\delta x^k(\XX)} \frac{\delta D}{\delta M_k(\XX)}
	= \int\dx s(\xx)\partial_j \frac{\delta C}{\delta s(\xx)} \frac{\delta D}{\delta m_j(\xx)}.
\end{equation}
So far we have recovered the $\{C,D\}^{(FM)}$ part of the bracket (the antisymmetric part is obtained as negative of the same terms with $C$ and $D$ swapped).

The part dependent on the Eulerian deformation gradient $\FF(\xx)$ is calculated similarly as follows.
\begin{align}
	&\int\dX \int\dx \frac{\delta C}{\delta \F{i}{I}(\xx)} \frac{\delta \F{i}{I}(\xx)}{\delta 
	x^k(\XX)} \frac{\delta D}{\delta M_k(\XX)}\nonumber\\
	&\qquad=\int\dX \int\dx \frac{\delta C}{\delta \F{i}{I}(\xx)}\left[
	- \frac{\partial \delta(\XX-\XX(\xx))}{\partial X^I}\delta^i_k
	-\partial_k \F{i}{I}(\xx)\delta(\XX-\XX(\xx))\right]\frac{\delta D}{\delta 
	m_k(\xx)}\Big|_{\xx(\XX)}\nonumber\\
&\qquad=\int\dx  \frac{\delta C}{\delta \F{i}{I}(\xx)}\left[
	\int\dX \delta(\XX-\XX(\xx)) \frac{\partial}{\partial X^I}\frac{\delta D}{\delta m_i(\xx)}\Big|_{\xx(\XX)}
	-\partial_k \F{i}{I}(\xx)\frac{\delta D}{\delta m_k(\xx)}\right]\nonumber\\
&\qquad=\int\dx  \frac{\delta C}{\delta \F{i}{I}(\xx)}\left[
	\frac{\partial x^j}{\partial X^I}\Big|_{\XX(\xx)}\frac{\partial}{\partial x^j}\frac{\delta D}{\delta m_i(\xx)}
	-\partial_k \F{i}{I}(\xx)\frac{\delta D}{\delta m_k(\xx)}\right]\nonumber\\
&\qquad=\int\dx  \frac{\delta C}{\delta \F{i}{I}(\xx)}\left(\F{j}{I}(\xx)\partial_j \frac{\delta 
D}{\delta m_i(\xx)}
	-\partial_k \F{i}{I}(\xx)\frac{\delta D}{\delta m_k(\xx)}\right),
\end{align}
which is the remaining part of bracket \eqref{eq.PB.Eu}. 

In summary, Eulerian Poisson bracket \eqref{eq.PB.Eu}, which expresses kinematics of fields $\rho(\xx)$, $\mm(\xx)$, $s(\xx)$ and $\FF(\xx)$, has been derived by reduction of the Lagrangian canonical Poisson bracket \eqref{eq.PB.L}, expressing kinematics of $\xx(\XX)$ and $\MM(\XX)$.

\section{From deformation gradient to the Left Cauchy-Green tensor}\label{sec.F-B}
Derivative of the left Cauchy-Green tensor $\BB(\xx)$ with respect to the Eulerian deformation gradient is
\begin{equation}
	\frac{\partial B^{ij}(\xx)}{\partial \F{k}{K}} = \delta^{\sI\sJ}(\delta^i_{\ k} \delta^\sK_\sI 
	\F{j}{J} +  
	\F{i}{I} \delta^j_{\ k} \delta^\sK_\sJ)
	= \delta^{\sK\sJ}\delta^i_k \F{j}{J} +  \delta^{\sI\sK} \F{i}{I} \delta^j_k.
\end{equation}
Derivative of a functional $C(\FF)$ then becomes
\begin{equation}
	\frac{\delta C}{\delta \F{k}{K}(\xx)} = 
	\frac{\delta C}{\delta B^{kj}(\xx)}\delta^{\sK\sJ} \F{j}{J} +  \frac{\delta C}{\delta 
	B^{ik}(\xx)}\delta^{\sI\sK} \F{i}{I},
\end{equation}
and after plugging this relation into bracket \eqref{eq.PB.Eu} we obtain bracket \eqref{eq.PB.B} easily.

\section{From deformation gradient to distortion}\label{sec.F-A}
The purpose of this section is to show more details on the reduction of bracket \eqref{eq.PB.Eu} to bracket \eqref{eq.PB.A}, expressing kinematics of distortion. The distortion is the inverse of the Eulerian deformation gradient $\FF(\xx)$, 
\begin{equation}
	\A{I}{i}(\xx) \F{j}{I}(\xx) = \delta^j_{\ i}.
\end{equation}
Taking derivative of this equality with respect to $\F{k}{K}(\xx)$ leads to
\begin{equation}
	\frac{\partial \A{I}{i}}{\partial \F{k}{K}} \F{j}{I} = -\A{K}{i} \delta^j_{\ k}.
\end{equation}
After multiplication by $\A{J}{j}$ we obtain
\begin{equation}\label{app.eq.dAdF}
	\frac{\partial \A{L}{l}}{\partial \F{j}{J}} = -\A{J}{l} \A{L}{j},
\end{equation}
from which it follows that 
\begin{equation}
	\frac{\delta C}{\delta \F{j}{J}} = \frac{\delta C}{\delta \A{L}{l}(\xx)} \frac{\partial 
	\A{L}{l}}{\partial \F{j}{J}} = -\frac{\delta C}{\delta \A{L}{l}} \A{J}{l} \A{L}{j}
\end{equation}
for arbitrary functional $C(\FF)$.

Plugging this last relation into bracket \eqref{eq.PB.Eu} immediately leads to bracket \eqref{eq.PB.A}.

\section{Additional mass transport}\label{sec.rho0}
In \ref{sec.L-E} energy depends on a constant Lagrangian field $\rho_0(\XX)$, apart from the state variables $\xx(\XX)$ and $\MM(\XX)$. It is possible to promote the field $\rho_0$ to a state variable and allow for its independent evolution. Let us thus start with a canonical couple $(\rho_0(\XX), \psi(\XX))$ equipped with the canonical Poisson bracket
\begin{equation}
\{F,G\}^{(can)} = \int\dX \left(\frac{\delta F}{\delta \rho_0} \frac{\delta G}{\delta \psi}-\frac{\delta G}{\delta \rho_0} \frac{\delta F}{\delta \psi}\right).
\end{equation}
Projecting to the gradient of $W_I(\XX) = \partial_{\XX}\psi(\XX)$, the canonical Poisson bracket becomes
\begin{equation}\label{eq.PB.rho0W}
\{F,G\}^{(\rho_0,\WW)} = -\int\dX \left(\frac{\delta F}{\delta \rho_0} \frac{\partial}{\partial X^I}\frac{\delta G}{\delta W_I}-\frac{\delta G}{\delta \rho_0} \frac{\partial}{\partial X^I}\frac{\delta F}{\delta W_I}\right).
\end{equation}
Assuming quadratic dependence of energy on $\WW$, this Poisson bracket expresses a wave equation for $\rho_0$. 

The goal now is to transform the evolution of $(\rho_0, \WW)$ into the Eulerian frame by transformation to $\rho(\xx)$ (as in Eqs. \eqref{eq.x.E}) and
\begin{equation}
w_i(\xx) = \frac{\partial X^I}{\partial x^i} W_I(\XX(\xx)).
\end{equation}
Derivative of an arbitrary functional $C$ with respect to $\rho_0$ becomes
\begin{equation}
\frac{\delta C}{\delta \rho_0(\XX)} = \int \dx \delta(\XX-\XX(\xx)) \frac{\delta C}{\delta \rho(\xx)} \det \frac{\partial \XX}{\partial \xx}
\end{equation}
and with respect to $\WW$ it becomes 
\begin{equation}
\frac{\delta C}{\delta W_I(\XX)} = \int \dx \delta(\XX-\XX(\xx)) \frac{\delta C}{\delta w_i(\xx)} \A{I}{i}(\xx).
\end{equation}
Derivative of an arbitrary functional $C$ with respect to the field $\xx$ then contains extra terms 
\begin{eqnarray}
\frac{\delta C}{\delta x^k(\XX)} &=& \mbox{Eq. \eqref{eq.Cx}}\\
&&+ \int \dx \frac{\delta F}{\delta w_j(\xx)} \left(w_k(\xx)\A{J}{j}(\xx)\frac{\partial \delta(\XX-\XX(\xx))}{\partial X^J} -\partial_k w_j(\xx)\delta(\XX-\XX(\xx))\right).\nonumber
\end{eqnarray}
Derivative of the functional with respect to $\MM$ is not changed.

Poisson brackets \eqref{eq.PB.L} and \eqref{eq.PB.rho0W} then become (see Eq. \eqref{eq.PB.A})
\begin{eqnarray}
\{F,G\}^{(\rho,\ww,\mm,s,\AA)} &=& \{F,G\}^{(A)} + \nonumber\\
&&+\int\dx \left(\partial_i F_\rho G_{w_i}-\partial_i G_\rho F_{w_i}\right)\nonumber\\
&&+\int\dx w_k \left(F_{w_j} \partial_j G_{m_k}-G_{w_j} \partial_j F_{m_k}\right)\nonumber\\
&&-\int\dx \partial_k w_j \left(F_{w_j} G_{m_k}-G_{w_j} F_{m_k}\right),
\end{eqnarray}
which expresses reversible evolution of fields $(\rho, \ww, \mm, s, \AA)$. The evolution equations implied by this bracket are
\begin{subequations}\label{eq.evo.rhow}
	\begin{eqnarray}
	\partial_t \rho &=& -\partial_i(\rho E_{m_i}-E_{w_i})\\
	\partial_t w_i &=& -\partial_i E_\rho + w_k \partial_i E_{m_k} -\partial_k w_i E_{m_k}\\
	\partial_t m_i &=& -\partial_j(m_i E_{m_j})-\rho\partial_i E_\rho - m_j \partial_i E_{m_j} -s 
	\partial_i E_s - \A{L}{l}\partial_i E_{\A{L}{l}} - w_j \partial_i E_{w_j}\nonumber\\
		&&+\partial_i(\A{L}{l} E_{\A{L}{l}}) - \partial_l(\A{L}{i} E_{\A{L}{l}})\nonumber\\
&&+\partial_j(w_i E_{w_j}) +\partial_i (w_j E_{w_j})\\
	\partial_t s &=& -\partial_i(s E_{m_i})\\
		\partial_t \A{L}{l} &=& -\partial_l (\A{L}{i} E_{m_i}) + (\partial_l \A{L}{i} - \partial_i 
		\A{L}{l}) 
		E_{m_i},
	\end{eqnarray}
\end{subequations}
which are the SHTC equations including the extra mass flux, see e.g. \cite{Peshkov-Grmela}.

\section{Rigorous generalization or the gauge invariance}\label{sec.rigorous}

Theorem 7.15 from \cite{olver2000applications} reveals an equivalence between a 
conservation law and a symmetry but on top of that it provides a connection 
between the conserved quantity $\Gfunc$ and a particular symmetry of the 
system. This symmetry corresponds to a Hamiltonian vector field $\vv_\Gfunc$ 
with characteristic (components) $\QQ=\{\qq,\Gfunc\}$ (in our notation) and 
where state variables are denoted as $\qq$. Now, a Hamiltonian vector field is 
a special type of evolutionary vector field (having non-zero components only 
those corresponding to the state variables) such that its prolongation is equal 
to the Poisson bracket in the following sense 
\begin{equation}
	\mathrm{pr} ~ \vv_\QQ(\Ffunc) \stackrel{\mathrm{def}}{=} \{\Ffunc,\QQ\} 
	\quad\forall\,\Ffunc(t,\xx,\qq).
\end{equation}
Note that it 
can be shown that for any Poisson bracket and any functional $\Ffunc$, such an 
evolutionary vector field exists. In particular, given $\vv = \tau \partial_t + 
\xi^i 
\partial_{x^i} + \phi^\alpha \partial_{q^\alpha}$,
$\tau$, $\xi^i$ and $\phi^\alpha$ being components of the vector field,
then a 
corresponding evolutionary vector field (which form is particularly convenient 
for calculating prolongations) is 
\begin{equation}
	\vv_Q =  Q^\alpha(t,\xx,\qq,\nabla \qq) \partial_{q^\alpha}
\end{equation}
where $Q^\alpha = \phi^\alpha - \xi^i \partial_{x^i} q^\alpha - \tau \partial_t 
q^\alpha$.
A vector field $\vv$ is a 
symmetry of 
a system if and only if its 
evolutionary representative vector field $\vv_Q$ is a symmetry of the system. 

First, we make few observations. 
\begin{description}
    \item[finite dimension] For a canonical Poisson bracket we have that 
    $$\vv=\vv_\QQ = \mathrm{pr}~ \vv = \mathrm{pr}~ \vv_\QQ,$$ as all the 
    derivatives are only with respect to state variables and hence 
    $$\vv=\mathrm{pr}~\vv_\Gfunc = \{\mathbf{\qq},\Gfunc\}$$ defines an 
    infinitesimal transformation of dependent (state) variables 
    $$\mathbf{\bar{\qq}} = \mathbf{\qq} + \varepsilon \{\mathbf{\qq},\Gfunc\}$$
    which is actually a symmetry of the system. Note that this observation extends to any finite dimensional system due to Darboux' theorem stating that any Poisson bracket in finite dimensions can be rewritten locally into a canonical form.
    \item[algebraic Poisson bracket] By repeating the same arguments as in the 
    previous point we can similarly show that for an algebraic Poisson bracket 
    (now independent of dimension) a transformation of dependent (state) 
    variables $$\mathbf{\bar{\qq}} = \mathbf{\qq} + \varepsilon 
    \{\mathbf{\qq},\Gfunc\}$$
    is a symmetry of the system.
\end{description}

We shall now proceed to the general case where we already know that symmetry 
exists (the system is invariant to this transformation) once we know there is a 
conserved quantity $\Gfunc$. The aim is to find this transformation explicitly. 
From the theorem 
(7.15 in \cite{olver2000applications}) we know that 
$\{q^\alpha,\Gfunc\}=Q^\alpha$ defines a characteristic 
$Q^\alpha$ of the evolutionary vector field and hence it has to be of the 
following form 
\begin{equation}
	Q^\alpha = \phi^\alpha(t,\mathbf{x},\mathbf{q}) - \xi^i(t,\mathbf{x},\mathbf{q}) \partial_{x^i} q^\alpha - \tau(t,\mathbf{x},\mathbf{q}) \partial_t q^\alpha,
\end{equation}
which then determines a transformation (via the corresponding vector field $\vv = \xi^i \partial_{x^i} + \tau \partial_t + \phi^\alpha \partial_{q^\alpha}$)
\begin{subequations}\label{eq.Ol.trafo}
\begin{align}
    \bar{t}    &= t + \varepsilon \tau(t,\xx,\qq) ,\\
    \bar{x}^i  &= x^i + \varepsilon \xi^i(t,\xx,\qq),\\
    \bar{q}^\alpha  &= q^\alpha + \varepsilon \phi^\alpha(t,\xx,\qq).
\end{align}
\end{subequations}
The two above special cases can now be easily understood as well as they correspond to the special situation when $\xi^i=0,~\tau=0$. 

Hence we showed the following statement:
\begin{theorem}
	Any Hamiltonian system with a conserved quantity $\Gfunc$ is invariant to 
	transformation \eqref{eq.Ol.trafo}
with $\{q^\alpha,\Gfunc\}= \phi^\alpha(t,\mathbf{x},\mathbf{q}) - 
\xi^i(t,\mathbf{x},\mathbf{q}) \partial_{x^i} q^\alpha - 
\tau(t,\mathbf{x},\mathbf{q}) \partial_t q^\alpha$, providing identification of 
$\phi^\alpha,~\xi^\alpha$ and $\tau$ in a unique way, cf. \cite{olver2000applications}.
\end{theorem}

Note that transformations \eqref{eq.barq} and \eqref{eq.Ol.trafo} are 
compatible as infinitesimal transformations, since the former can be seen as 
Taylor series to the first order of the latter. Therefore, the former can be 
seen as transformations in a broader sense (allowing also for differential 
operators).

Let us illustrate the above observations on a few examples.

\paragraph{Hamiltonian mechanics of particles.} State variables are position 
and 
momentum, $\qq=(\rr,\pp)$, while the only independent variable is time $t$, and 
the Poisson bracket is in a canonical form. In this case we know that the 
relation between conserved quantity $\Gfunc$ and invariant transformation is 
particularly simple: $\bar{\rr} = \rr + \varepsilon \{\rr,\Gfunc\},~ \bar{\pp} 
= 
\pp + \varepsilon \{\pp,\Gfunc\}$, as in Eq. \eqref{eq.barq}.

From conservation of particle momentum, $\Gfunc=c^i p_i$ (for an arbitrary fixed vector $\cc$), we 
immediately get 
invariance to translations $\bar{r}^j = r^j + \varepsilon c^j$, 
$\bar{\pp} = 
\pp$. Conservation of 
angular momentum $\Gfunc = \cc \cdot (\rr\times \pp)$ 
yields invariance to rotations $\bar{\rr} = \rr + \varepsilon \cc \times \rr$. 
Finally, conservation of Galilean boost $\Gfunc = m \mathbf{r} - t \mathbf{p}$, $m$ being mass of the particle\footnote{not to be confused with continuum momentum density $\mm$},
entails Galilean invariance $\bar{\rr} = \rr + \varepsilon \{\rr,\Gfunc\}= \rr 
- t 
\Id ,~ \bar{\pp} = \pp + \varepsilon \{\pp,\Gfunc\} = \pp - m\Id$.

\paragraph{Korteweg-de Vries.} equation reads
\begin{equation}
    u_t = u_{xxx}+u u_x = \{u,\mathcal{H}\}, 
\end{equation}
where
\begin{equation*}
    \mathcal{H}[u] = \int -\frac 1 2 u_x^2+ \frac 1 6 u^3 \mathrm{d} x,\quad \{A,B\} = \int \delta A \frac{\mathrm{d}}{\mathrm{d}x} \delta B \mathrm{d}x.
\end{equation*}
Conservation of the following functionals
\begin{equation*}
    P_1 = \int \frac 1 2 u^2 \mathrm{d} x,\quad P_2 = \int (\frac 1 6 u^3 - \frac 1 2 u_x^2) \mathrm{d} x, \quad P_3 = \int(xu+\frac 1 2 t u^2) \mathrm{d}x,
\end{equation*}
corresponds to characteristics
\begin{equation*}
    Q_1 = \{u,P_1\} = u_x,\quad Q_2 = \{u,P_2\} = u_{xxx}+ u u_x = u_t, \quad Q_3 = \{u,P_3\} = 1 + t u_x,
\end{equation*}
and hence, to invariants of the problem
\begin{equation*}
	\bar{t} = t + \varepsilon,
	\quad 
    \bar{x} = x + \varepsilon,
    \quad
    \bar{u}(x,t) = u(x-\varepsilon t,t) + \varepsilon. 
\end{equation*}

\paragraph{Lagrangian frame of fluid mechanics.}
First, let us choose the Lagrangian frame, bracket \eqref{eq.PB.L}, and 
$\Gfunc=\int\dX M_j(\XX)$ equal to the $ j $-th component of the total momentum. Using energy 
\eqref{eq.Ene.L}, the Poisson bracket of $\Gfunc$ and energy is
\begin{equation}
    \{\Gfunc, E\}^{(L)} = -\int\dX E_{x^j} \delta^j_{\ k} = \int\dX 
    \frac{\partial}{\partial X^\sJ} \left(\rho_0(\XX) \frac{\partial W}{\partial 
    \frac{\partial x^k}{\partial X^\sJ}}\right) = 0,
\end{equation}
which means that total the momentum is conserved. Infinitesimal transformations of the state 
variables $\xx(\XX)$ and $\MM(\XX)$ are (as above)
\begin{subequations}
\begin{eqnarray}
    \bar{x}^i(\XX) &=& x^i(\XX) + \eps \{x^i(\XX),\Gfunc\}^{(L)} = x^i(\XX) 
    +\eps \delta^i_{\ k}\\
    \bar{M}_i(\XX) &=& M_i(\XX) + \eps \{M_i(\XX),\Gfunc\}^{(L)} = M_i(\XX).
\end{eqnarray}
\end{subequations}
The first equation expresses infinitesimal translation in the $k$-direction and 
hence, the evolution equations are invariant with respect to infinitesimal 
translations, as it follows directly from the conservation of the total momentum 
$\Gfunc$.

\paragraph{Eulerian frame of fluid mechanics.}
In the Eulerian setting, let us choose the Poisson bracket for fluid mechanics \eqref{eq.PB.FM} and state variables $(\rho,\mm,s)$. 

The consequence of energy conservation is that $\{\rho,E\}$ has to be a characteristic of the 
evolutionary vector field. Indeed, we have
\begin{equation*}
    \{\rho,E\}=Q_\rho= \partial_t \rho,
\end{equation*}
which yields ($\phi_\rho=0,~\xi^i=0,~\tau=-1$) invariance of the density field 
to the transformation $\tilde{t} = t+ \varepsilon$. In the same manner all 
other state variables are invariant translation in time.

Further let us inspect the implications of conservation of the total momentum 
$\Gfunc=\int\dx m_j(\xx)$. Calculating the symmetry 
\begin{equation*}
    \{\rho,\Gfunc\} = \{\rho,m_j\} = \int \dx _b \rho(\xx_b) \frac{\partial 
    \delta(\xx_a-\xx_b)}{\partial x_b^j} = -\partial_j \rho
\end{equation*}
reveals that density is invariant to translations, $\tilde{x}^i = x^i + 
\varepsilon$. Similarly, from
\begin{equation*}
    \{m_i,\Gfunc\} = \{m_i,m_j\} = \int \dx_b m_i(\xx_b) \frac{\partial 
    \delta(\xx_a-\xx_b)}{\partial x_b^j} - m_j(\xx_a) \frac{\partial 
    \delta(\xx_a-\xx_b)}{\partial x_a^j} = -\partial_j m_i,
\end{equation*}
we have the same observation that the momentum is invariant to translations. Then, from
\begin{equation*}
    \{s,\Gfunc\} = \{s,m_j\} = \int \dx_b s(\xx_b) \frac{\partial 
    \delta(\xx_a-\xx_b)}{\partial x_b^j} = -\partial_j s,
\end{equation*}
we can conclude that fluid mechanics is invariant to translations as we know that the total 
momentum is conserved.

Finally, let us choose $\Gfunc$ as the overall Galilean booster, see e.g. 
\cite{extra-mass-flux}, 
\begin{equation}\label{eq.booster}
    \Gfunc_i = \int\dx \underbrace{(\rho(\xx) x_i - t 
    m_i(\xx))}_{g_i(\xx)}.
\end{equation}
The booster is indeed conserved, since
\begin{eqnarray}
\dot{\Gfunc}_i &=& \int \dx \partial_t g_i= \nonumber\\
&=& -\int\dx m_i+ \int\dx \left(\rho E_{m_i} + t(\rho\partial_i E_\rho +m_j \partial_i E_{m_j} +s\partial_i E_s)\right)\nonumber\\
&=& -\int\dx (m_i - \rho E_{m_i}) + t \int\dx \partial_i p = 0
\end{eqnarray}
for energy 
\begin{equation}
E = \int\dx\left(\frac{\mm^2}{2\rho} + \varepsilon(\rho,s)\right),
\end{equation}
$p$ being the pressure, see e.g. \cite{PKG}.
The fluid fields transform to 
\begin{subequations}
\begin{eqnarray}
    \bar{\rho} &=& \rho +\eps\{\rho,\Gfunc_i\}^{(FM)} = \rho +\eps 
    t\partial_i \rho\\
    \bar{m_j} &=& \rho +\eps\{m_j,\Gfunc_i\}^{(FM)} = m_j + \eps 
    t\partial_i m_j - \eps \rho \delta_{ji}\\
    \bar{s} &=& s +\eps\{s,\Gfunc_i\}^{(FM)} = s +\eps t\partial_i s,
\end{eqnarray}
\end{subequations}
which corresponds to infinitesimal transformation
\begin{eqnarray}
    \tau&=0,\\
    \xi^i&=-\delta_{ij} t,\\
    \phi_\rho &= 0,\\
    \phi_{m_i} &= -\rho \delta_{ij},\\
    \phi_s &=0,
\end{eqnarray}
representing Galilean transformation. Hence, we showed that Galilean boost is conserved and as a 
result the dynamics of the system is invariant to Galilean transformation.

Similarly, for the extended bracket \eqref{eq.PB.A}, which expresses kinematics of fields $\rho$, 
$\mm$, $s$ and $\AA$, the Galilean booster \eqref{eq.booster} is also conserved as it is followed 
from a similar calculation as above. The infinitesimal Galilean transformation then reads
\begin{subequations}
\begin{eqnarray}
    \bar{\rho} &=& \rho +\eps\{\rho,\Gfunc_k\}^{(A)} = \rho +\eps t\partial_k 
    \rho\\
    \bar{m_i} &=& \rho +\eps\{m_i,\Gfunc_k\}^{(A)} = m_i + \eps t\partial_k m_i - 
    \eps \rho \delta_{ik}\\
    \bar{s} &=& s +\eps\{s,\Gfunc_k\}^{(A)} = s +\eps t\partial_k s\\
    \bar{A}^\sJ_{\ j} &=& \A{J}{j} +\eps\{\A{J}{j},\Gfunc_k\}^{(A)} =  \A{J}{i} + 
    \eps 
    t 
    \partial_k \A{J}{j},
\end{eqnarray}
and  evolution equations \eqref{eq.evo.A} are transformed in the same way: they are Galilean 
invariant.
\end{subequations}

\section{Proof of the Godunov-Boillat theorem}\label{sec.GBproof}
Let us briefly recall the proof the Godunov-Boillat theorem.
Let the state variables be denoted by $q^\alpha$, and assume that their evolution equations are in conservative form,
\begin{equation}
\partial_t q^\alpha = -\partial_i J^{\alpha i}(\qq).
\end{equation}
Assume moreover, that total energy density is a convex function of the state variables $e(\qq)$, and that there is an extra conservation law implied by the evolution equations for $\qq$, 
\begin{eqnarray}
\partial_t e = -\partial_i J^{ei} = -\frac{\partial J^{ei}}{\partial q^\alpha} \partial_i q^\alpha 
= -\frac{\partial e}{\partial q^\alpha} \frac{\partial J^{\alpha i}}{\partial q^\beta}\partial_i q^\beta,
\end{eqnarray}
i.e. satisfying the compatibility condition
\begin{equation}\label{eq.GB.comp}
\frac{\partial J^{ei}}{\partial q^\beta} = \frac{\partial e}{\partial q^\alpha} \frac{\partial J^{\alpha i}}{\partial q^\beta}.
\end{equation}

Variables $\pp$ conjugate to $\qq$ w.r.t. energy $e$ are defined by Legendre transformation, Eqs. \eqref{eq.qp}, and the Legendre transformation of energy, convexity of which follows from convexity of energy, is denoted by $L(\pp)$.
The conjugate energy flux 
\begin{equation}
J^{\dagger i}(\pp) = -J^{ei}(\qq(\pp))+\left(\frac{\partial e}{\partial q^\alpha} J^{\alpha i}\right)\Big|_{\qq(\pp)}
\end{equation}
has simple derivatives w.r.t. the conjugate variables,
\begin{eqnarray}
\frac{\partial J^{\dagger i}}{\partial p_\beta} &= &
 \left(-\frac{\partial J^{ei}}{\partial q^\alpha}\frac{\partial q^\alpha}{\partial p_\beta} 
 +\frac{\partial^2 e}{\partial q^\alpha\partial q^\gamma}\frac{\partial q^\gamma}{\partial p_\beta} J^{\alpha i}
 +\frac{\partial e}{\partial q^\alpha} \frac{\partial J^{\alpha i}}{\partial q^\gamma}\frac{\partial q^\gamma}{\partial p_\beta}\right)\Big|_{\qq(\pp)}\nonumber\\
 &\stackrel{\mbox{Eq. \eqref{eq.GB.comp}}}{=}&
 \frac{\partial^2 e}{\partial q^\alpha\partial q^\gamma}\Big|_{\qq(\pp)}\frac{\partial q^\gamma}{\partial p_\beta} J^{\alpha i}\Big|_{\qq(\pp)}
 =\frac{\partial^2 e}{\partial q^\alpha\partial q^\gamma}\Big|_{\qq(\pp)}\frac{\partial^2 L}{\partial p_\gamma \partial p_\beta} J^{\alpha i}\Big|_{\qq(\pp)}\nonumber\\
 &=& J^{\beta i}\Big|_{\qq(\pp)}
\end{eqnarray}
because the Hessians of $e$ and $L$ are inverse of each other.

Evolution equations for $\qq$ can be then rewritten as
\begin{equation}
\frac{\partial^2 L}{\partial p_\alpha \partial p_\beta}\partial_t p_\beta = \frac{\partial^2 J^{\dagger i}}{\partial p_\alpha \partial p_\beta} \partial_i p_\beta,
\end{equation}
which is a system of symmetric hyperbolic PDEs of first order (the matrix in front of the time derivative being symmetric positive definite and the matrix in front of the spatial derivative symmetric).

\end{document}